\documentclass[preprint,3p,authoryear]{elsarticle}

\journal{Planetary \& Space Science}

\usepackage{natbib}

\usepackage{amssymb}

\usepackage{doi}
\usepackage{textcomp}
\usepackage{hyperref}
\hypersetup{colorlinks=true, urlcolor=blue}

\begin{document}

\begin{frontmatter}



\title{The reflectance spectrum of Titan's surface at the Huygens landing site determined by the Descent Imager/Spectral Radiometer\tnoteref{label1}\tnoteref{label2}}
\tnotetext[label1]{\doi{10.1016/j.pss.2007.10.011}}
\tnotetext[label2]{\copyright 2017. This manuscript version is made available under the CC-BY-NC-ND 4.0 licence:\\ \url{https://creativecommons.org/licenses/by-nc-nd/4.0/}}


\author{S.E. Schr\"oder}
\author{H.U. Keller}

\address{Max-Planck-Institut f\"ur Sonnensystemforschung, Max-Planck-Strasse 2, 37191 Katlenburg-Lindau, Germany}

\begin{abstract}

The Descent Imager/Spectral Radiometer aboard the Huygens probe successfully acquired images and spectra of the surface of Titan. To counter the effects of haze and atmospheric methane absorption it carried a Surface Science Lamp to illuminate the surface just before landing. We reconstruct the reflectance spectrum of the landing site in the 500-1500~nm range from Downward Looking Visual and Infrared Spectrometer data that show evidence of lamp light. Our reconstruction is a followup to the analysis by \citet{T05}, who scaled their result to the ratio of the up- and down flux measured just before landing. Instead, we use the lamp flux from the calibration experiment, and find a significantly higher overall reflectance. We attribute this to a phase angle dependance, possibly representing the opposition surge commonly encountered on solar system bodies. The reconstruction in the visible wavelength range is greatly improved. Here, the reflectance spectrum features a red slope, consistent with the presence of organic material. We confirm the blue slope in the near-IR, featureless apart from a single shallow absorption feature at 1500~nm. We agree with Tomasko et al.\ that the evidence for water ice is inconclusive. By modeling of absorption bands we find a methane mixing ratio of 4.5$\pm$0.5\% just above the surface. There is no evidence for the presence of liquid methane, but the data do not rule out a wet soil at a depth of several centimeters.

\end{abstract}


\end{frontmatter}


\section{Introduction}
\label{sec:introduction}

The highly successful Huygens mission has provided us with a unique view of Titan's surface. During its descent the Descent Imager/Spectral Radiometer (DISR) imaged bright, rugged terrain incised with rivers, terminating in a dark lake along a well-defined coastline \citep{T05}. Post-landing images show the lake bed to be dry and littered with cobbles. Over the past few years, the Cassini orbiter has revealed Titan to be a dynamic world, with rivers, clouds, and land forms which are highly suggestive of cryovolcanism \citep{E05,P05}. While the once proposed global methane/ethane ocean \citep{S82,L83} has not materialized \citep{W05}, lakes were recently discovered in the northern hemisphere \citep{S07}. Noticeable is the clear distinction between relatively dark terrain distributed around the equator, covered by dunes \citep{L06}, and the brighter terrain at higher latitudes. Since Huygens descended over the boundary of these two types of terrain, just south of some dunes \citep{LE07,MPS07}, its observations may help us understand this dichotomy. Even though they provide only a snapshot at a single location, they do represent the necessary ground truth for Cassini.

Attempts to investigate Titan's surface are generally hampered by its thick atmosphere. Aerosols generated by the photodissociation of methane in the stratosphere virtually obscure the surface at ultraviolet and visible wavelengths, while methane predominantly absorbs light in the near-infrared, except in certain narrow wavelength intervals. Ground based \citep{C95,G03,L04} and Cassini \citep{McC06} spectroscopic observations of the surface in these methane windows are usually interpreted in terms of water ice and a second, dark material. An important candidate for the latter are tholins, the solid end-product of the photodissociation of methane \citep{C79,C91,B06}. In support, Cassini's radar \citep{E05} and Huygens' GCMS \citep{N05} have found evidence for the presence of organic material on the surface. But so far, tholins have not unequivocally been detected. And because the near-IR methane windows roughly coincide with the water absorption line cores, these water lines cannot be resolved from outside the atmosphere and their existence has thus far been inferred. This is where DISR's Downward Looking Visual (DLVS) and Infrared (DLIS) Spectrometers enter the discussion. By actively illuminating the surface with its Surface Science Lamp (SSL) DISR could record a complete surface reflectance spectrum, albeit in a limited wavelength range. Complementary to ground based and Cassini observations, the DISR reflectance spectrum may better constrain the surface composition at the Huygens landing site, which will certainly lead to further understanding of the atmospheric methane cycle \citep{T06,A06}.

\citet{T05} have made the first reconstruction of the surface reflectance spectrum from DISR measurements. They scaled the last DLIS spectrum acquired before landing, which shows a contribution of reflected lamp light, to the ratio of the up- and downward flux determined from spectra uncontaminated by the lamp. Their treatment of the DLVS part of reflectance spectrum was preliminary, and led to significant artifacts. Their analysis leaves enough questions open to merit further investigation. For example, what do we find for the absolute reflectance if we take the lamp flux from the original calibration experiments? What is the shape of the visual reflectance spectrum as derived from the DLVS? What is the sensitivity of the reflectance reconstruction to the choice of background spectrum? How do newly available methane absorption coefficients and the revised altitude scale of Huygens' descent affect the analysis? And finally, what is the significance of the inconsistency between the reflectance derived from pre- and post-landing spectra that is seen around 1500~nm in Fig.~15a in \citet{T05}? This paper addresses these questions. In addition, we look for changes in spectra acquired on the surface for the period of more than an hour in which probe telemetry was received.

It will not have escaped the reader's attention that we make frequent use of acronyms; the most important ones are listed in Table~\ref{tab:acronyms}. Some details of the analysis have been omitted from this paper; they are provided by \citet{SES07}.

\begin{table}
\caption{Acronyms used throughout this paper.}
\vspace{5mm}
\centering
\begin{tabular}{|ll|}
\hline \hline
{\bf DISR} & Descent Imager / Spectral Radiometer \\
{\bf DLIS} & Downward Looking Infrared Spectrometer \\
{\bf DLVS} & Downward Looking Visual Spectrometer \\
{\bf HRI} & High Resolution Imager \\
{\bf MNS} & Medium Near Surface \\
{\bf MRI} & Medium Resolution Imager \\
{\bf SLI} & Side Looking Imager \\
{\bf SSL} & Surface Science Lamp \\
{\bf ULIS} & Upward Looking Infrared Spectrometer \\
{\bf VLNS} & Very Low Near Surface \\
\hline \hline
\end{tabular}
\label{tab:acronyms}
\end{table}

\section{Observations}
\label{sec:observations}

\subsection{DISR operational modes}

Whereas the DLIS had its own linear array of sensors, the DLVS shared the CCD with the DISR imagers, occupying an area of 20~columns by 200~rows. DISR switched between different modes of operation during the descent to maximize the science return. Generally, adjacent columns of the DLVS section on the CCD were summed, and the resulting 10 columns were returned. We refer to this DLVS mode of operation as the 10-column mode. The full set of 20~columns was only returned during acquisition of the two spectrophotometric maps, which we analyze in a separate paper \citep{MPS07}. Regular DLIS exposures were summed on board over multiple rotations. This paper does not deal with regular spectra, but only with those acquired in the Very Low Near Surface (VLNS) and subsequent surface modes. At 210~m altitude (45~seconds before landing) DISR switched to the VLNS mode, in which it recorded DLVS/DLIS spectra and Violet Photometer observations only. In this mode the DLVS returned only two out of ten summed columns (4+5 and 6+7), a mode of operation known as the 2-column mode. In contrast to its regular mode of operation the DLIS now recorded brief, single exposures, and acquired dark exposures (with shutter closed) separately. Huygens landed at mission time 8869.77 seconds \citep{Z05}. After 48 seconds on the surface DISR switched to a surface mode in which it once again acquired images and DLVS spectra in the 10-column mode. As a safeguard (the probe's software might have been wrong about having landed!), the DLIS returned to the summing mode even though it remained fixated at the same spot on the surface. After landing DISR was active and transmitted data for more than an hour. The SSL was switched on 2~minutes and 16~seconds before landing (at 620~m altitude), and was left on for the remainder of the mission.

\subsection{Selected spectra}

DISR assigned a unique sequential number to each spectrum it recorded, which we print bold throughout this paper. We reconstruct the surface reflectance from the handful of (VLNS) spectra before landing which show evidence of lamp light. For the DLVS these are spectra {\bf 785} and {\bf 786}, acquired at 16 and 8~m above the surface, respectively. As background spectra we use two spectra acquired shortly before ({\bf 772} and {\bf 779}). The details of these spectra are listed in Table~\ref{tab:dlvs_VLNS}. The tabulated altitude and attitude are from the reconstruction by \citet{K07}, with exceptions noted. The altitude of the spectrometer windows after landing is 46~cm, with the probe azimuth angle estimated to be $257.0^\circ$. The first three spectra acquired after landing are partly overexposed. We select the first correctly exposed post-landing spectrum ({\bf 791}) and three later spectra (including the last one returned) for additional analysis (Table~\ref{tab:dlvs_surface}). Two DLIS spectra which show clear evidence for lamp light are {\bf 206} and {\bf 210}, acquired at 55 and 25~m above the surface, respectively. As background spectra we use two spectra acquired shortly before ({\bf 199} and {\bf 202}); details can be found in Table~\ref{tab:dlis_VLNS}. After landing, spectra acquired prior to {\bf 249} are all severely overexposed. We select three correctly exposed post-landing spectra (including the first and the last) for additional analysis (Table~\ref{tab:dlis_surface}).

\begin{table}
\centering \caption{Details of the pre-landing VLNS (2-column mode) DLVS spectra used in this paper. Mission time listed is halfway through the exposure. Azimuth is defined counterclockwise with respect to the east. The altitudes of DLVS {\bf 785} and {\bf 786} were calculated from the time of landing (mission time 8869.77 s), assuming an impact velocity of 4.60 m s$^{-1}$ as determined by the Surface Science Package \citep{Z05}. The second column refers to the DISR cycle of operations, and the last column to the solar phase angle of the observation.} \vspace{5mm}
\begin{tabular}{|rrccccc|}
\hline \hline
\# & cycle & mission & exposure & altitude & azimuth & phase \\
 & & time (s) & time (s) & (m) & angle ($^\circ$) & angle ($^\circ$) \\
\hline
{\bf 772} &     95 &      8844.01 &      1.36 & 116     &  51.9 & 44.7 \\
{\bf 779} &    102 &      8855.99 &      1.36 &  62     & 336.1 & 54.0 \\
{\bf 785} &    108 &      8866.27 &      1.36 &  16.1   & 276.4 & 45.5 \\
{\bf 786} &    109 &      8867.98 &      1.36 &   8.2   & 267.0 & 43.0 \\
\hline \hline
\end{tabular}
\label{tab:dlvs_VLNS}
\end{table}

\begin{table}
\centering
\caption{Details of the post-landing DLVS spectra used in this paper. Mission time listed is halfway through the exposure. DLVS {\bf 791} is a 2-column (VLNS) mode spectrum, the others are 10-column mode spectra.} \vspace{5mm}
\begin{tabular}{|rrcccrrcc|}
\hline \hline
\# & cycle & mission & exposure & & \# & cycle & mission & exposure \\
 & & time (s) & time (s) & &  & & time (s) & time (s) \\
\hline
{\bf 791} & 114 &  8876.41 & 0.33 & &  {\bf 950} & 158 & 10961.60 & 0.36 \\
{\bf 821} & 142 &  8931.13 & 0.30 & & {\bf 1077} & 165 & 13005.36 & 0.36 \\
\hline \hline
\end{tabular}
\label{tab:dlvs_surface}
\end{table}

\begin{table}
\centering \caption{Details of the pre-landing VLNS DLIS spectra used in this paper. During an operation the instrument acquired a single exposure with shutter open (`sample'), and none with shutter closed. Mission time listed is halfway through sampling. Azimuth is defined counterclockwise with respect to the east. The altitude of {\bf 210} was calculated from the time of landing as in Table~\ref{tab:dlvs_VLNS}.} \vspace{5mm}
\begin{tabular}{|rrcccccc|}
\hline \hline
\# & cycle & mission & sampling & operation & altitude & azimuth & phase \\
 & & time (s) & time (s) & time (s) & (m) & angle ($^\circ$) & angle ($^\circ$) \\
\hline
{\bf 199} &     96 &     8845.55 &  1.00   & 1.02   & 109     &  42.2 & 45.5 \\
{\bf 202} &     99 &     8850.68 &  1.00   & 1.02   &  86     &   9.4 & 52.6 \\
{\bf 206} &    103 &     8857.53 &  1.00   & 1.02   &  55     & 326.8 & 55,4 \\
{\bf 210} &    107 &     8864.38 &  1.00   & 1.02   &  24.8   & 286.9 & 51.1 \\
\hline \hline
\end{tabular}
\label{tab:dlis_VLNS}
\end{table}

\begin{table}
\centering \caption{Details of the post-landing DLIS spectra used in this paper. During an operation the instrument acquired multiple exposures with shutter open (`samples') and closed, which were summed on board. Mission time listed is halfway through the operation. For spectra {\bf 261} and {\bf 268} the sampling time of two of the eight regions was half the value in the table (see Sec.~\ref{subsec:spec_cal}).} \vspace{5mm}
\begin{tabular}{|rrcccc|}
\hline \hline
\# & cycle & mission & sampling & samples & operation \\
 & & time (s) & time (ms) & & time (s) \\
\hline
{\bf 249} & 146 &     8960.23 & 16.13 &   54 & 2.05 \\
{\bf 261} & 158 &    10942.72 & 16.13 & 2144 & 70.71 \\
{\bf 268} & 165 &    12901.64 &  8.06 & 3680 & 71.48 \\
\hline \hline
\end{tabular}
\label{tab:dlis_surface}
\end{table}

\subsection{Context}

\begin{figure}
\centering
\includegraphics[width=\textwidth,angle=0]{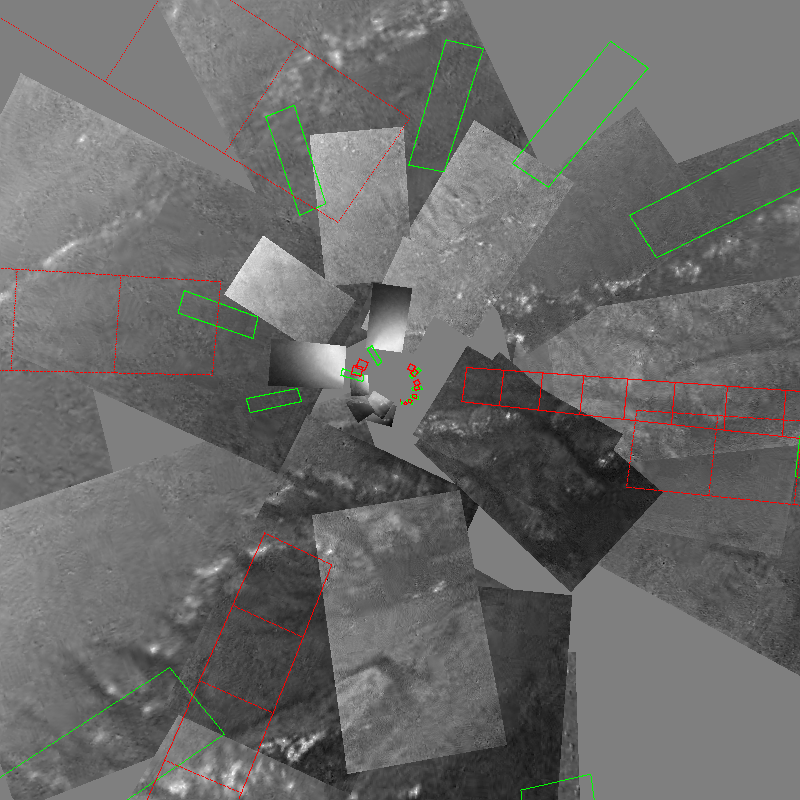}
\caption{The area around the Huygens landing site in Gnomonic projection (1.4 $\times$ 1.4~km). The landing site itself is in the center. The MNS and pre-landing VLNS mode DLVS (red) and DLIS (green) footprints are overlayed. In reality the DLIS MNS footprints are wider than shown here. All available high-res images (HRI {\bf 651}-{\bf 721}, MRI {\bf 664}-{\bf 700}) are shown. Terrain for which only low-res images are available is left gray. The last two full HRI images and the four half ones (HRI {\bf 711}-{\bf 721}) show internally scattered lamp light at the bottom. North is up, and east is to the right.}
\label{fig:MNS&VLNS_projection}
\end{figure}

\begin{figure}
\centering
\includegraphics[height=8cm,angle=0]{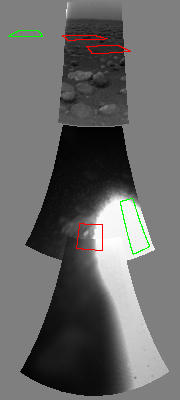}
\includegraphics[height=8cm,angle=0]{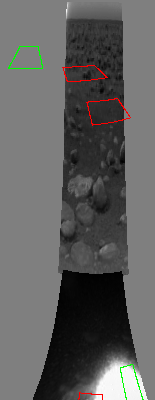}
\caption{The view from Huygens after landing in Mercator projection (7$^\circ$-100$^\circ$ nadir angle) from the probe's (MRI altitude 46 cm, {\bf left}) and a human's (altitude 1.7 m, {\bf right}) perspective. The last two DLVS and last DLIS pre-landing footprints are overlayed in red and green, respectively, and the post-landing footprints are visible at the bottom overlaying the (overexposed) lamp reflection spot. The intensity in the SLI image was scaled differently from the MRI and HRI for display purposes (images: HRI {\bf 1211}, MRI {\bf 1020}, SLI {\bf 742}).}
\label{fig:surface_projection}
\end{figure}

Figure~\ref{fig:MNS&VLNS_projection} shows the footprints of the spectra acquired in the last stage of the descent projected on high resolution images using the attitude and trajectory reconstruction of \citet{K07}. Shown is the small (1.4 $\times$ 1.4~km) area surrounding the landing site, which is located in the lake bed, approximately 3.5~km south of the coastline visible in the DISR mosaic \citep{T05,MPS07}. The pre-landing VLNS spectra are the small group of 11 single-footprint DLVS (red) and 7 DLIS (green) spectra in the center of the panorama, and were all acquired within a single probe rotation. Also shown are footprints of 10-column mode DLVS and single-exposure DLIS spectra acquired in the preceding Medium Near Surface (MNS) mode. Unfortunately, the area directly surrounding Huygens' landing site was imaged at very poor resolution; the last image to show the landing site itself (HRI {\bf 384}) was recorded at an altitude of 20~km. Consequently, none of the terrain covered by the VLNS footprints has been imaged at a resolution of better than circa 100~m. Most likely Huygens landed in the featureless gray terrain that dominates the area, but we cannot exclude that it landed on an extension of the relatively bright ridge that runs diagonally through Fig.~\ref{fig:MNS&VLNS_projection} from center right to bottom left. The footprints of the last two pre-landing DLVS spectra may have been captured by the surface SLI image. Figure~\ref{fig:surface_projection}, albeit not necessarily completely accurate in its positioning of the footprints, shows these to be located only a few meters from the probe. Hence, it is reasonable to assume that reflectance curves reconstructed from these spectra are representative for the terrain visible in the surface images.

\section{Methods}
\label{sec:methods}

\subsection{Surface reflectance reconstruction}
\label{subsec:refl_rec}

We reconstruct the surface reflectance from spectra which show evidence for lamp light by subtracting a background spectrum, and dividing the result by the lamp spectrum. The background spectrum is that which would have been observed if the lamp had been turned off, and needs to be constructed or selected. We approximate the true background by selecting a spectrum recorded closely in time with, and at nearly the same solar phase angle as the spectrum of interest. We vary its intensity to account for possible surface brightness variations. The criterion for judging the quality of a background spectrum is the smoothness of the final reflectance spectrum. That is, the shape and depth of the methane absorption bands, that appear due to the intervening atmosphere, should conform to reasonable assumptions. We assess this by modeling the methane absorption with the \citet{K98} coefficients below 1050~nm, and the \citet{I06} coefficients above.

To retrieve the reflectance in absolute units we need the SSL flux at Titan's surface. We find it by scaling the flux measured by the DISR\#3 flight model prior to launch by $1/d^2$, with $d$ the distance from the spectrometer window to the center of the footprint on the surface. This scaling is valid as long as parallax effects can be ignored (i.e.\ $d$ is large enough), and tip and tilt of the probe are zero. The spectrometer windows are located a few centimeters on opposite sides of the lamp on the DISR sensor head (DLIS: 3.2~cm, DLVS: 6.8~cm), leading to parallax effects close to the camera. We calculated that parallax can be safely ignored for the DLIS beyond 4~m, but that it makes the lamp flux as observed by the DLVS decrease slightly more with distance than expected from the $1/d^2$ scaling. This effect of the DLVS footprint `drifting away' from the lamp beam may lead to a lower reflectance as reconstructed from {\bf 785} (we did not correct for this). Swinging of the probe does not affect the reconstruction significantly; tilt values of less than a degree are predicted by \citet{K07}. Then we can estimate $d$ accurately from the altitude $h$ and the nadir angle of the spectrometer $\theta_{\rm Sp}$ if the surface is flat, as it appears to be in the post-landing SLI images (Fig.~\ref{fig:surface_projection}).

We reconstruct the reflectance as the radiance coefficient $r_{\rm C}$, or the bidirectional reflectance of the surface relative to that of an identically illuminated Lambert surface \citep{H81}, also known as ``$I/F$''\footnote{Postscript: We wish to point out that this statement, as it appears in the published article, is incorrect. $I/F$ is actually identical to the radiance factor $r_{\rm F}$, or the bidirectional reflectance of the surface relative to that of a Lambert surface illuminated normally \citep{H81}. The two are related by $r_{\rm F} = r_{\rm C} \cos \iota$, with $\iota$ the incidence angle of the light from the SSL, which was $20^\circ$-$22^\circ$ for the cases we study here. Please be aware that in the remainder of this paper, wherever we refer to the $I/F$, this actually means $(I/F) / \cos \iota$. To get the true $I/F$, one could multiply the results shown by a factor $\cos 20^\circ = 0.94$.}. We would like to subtract from a spectrum $I_1$ with clear presence of reflected lamp light a background spectrum, i.e.\ one that would have been recorded if the lamp were off. As we do not have such an ideal background spectrum we approximate it by taking an earlier recorded spectrum $I_2$ with much less reflected lamp light. Both spectra contain reflected lamp light (at identical phase angle $\phi$) and reflected sunlight (at different phase angles $\phi^\prime$ and $\phi^{\prime\prime}$). Then
\begin{equation}
\pi (I_1 - I_2) = r_{\rm C,1}(\phi) F_{\rm L}(h_1) + r_{\rm C,1}(\phi^\prime) F_\odot - r_{\rm C,2}(\phi) F_{\rm L}(h_2) - r_{\rm C,2}(\phi^{\prime\prime}) F_\odot
\label{eq:ideal_reflectance}
\end{equation}
with $F_{\rm L}$ the lamp flux and $F_\odot$ the solar flux at the surface (in W m$^{-2}$ \textmu m$^{-1}$). We now make two simplifying assumptions. The first is that the radiance coefficient is spatially constant: $r_{\rm C,1} = r_{\rm C,2} = r_{\rm C}$. The second is that the solar phase angle of both spectra is the same: $\phi^\prime = \phi^{\prime\prime}$. This leads to
\begin{equation}
r_{\rm C}(\phi) = \frac{\pi (I_1 - I_2)}{F_{\rm L}(h_1) - F_{\rm L}(h_2)}.
\label{eq:reflectance}
\end{equation}
The first assumption is reasonable \citep[see][]{MPS07}, and about the second we have to make sure by carefully selecting the background spectrum (the directional reflectance of the surface will be discussed in more detail in an upcoming paper). What is left now is to determine the lamp flux on Titan's surface $F_{\rm L}(d) = F_{\rm SSL}(d) \cos\theta_{\rm SSL}$ from the flux $F_{\rm SSL}$ measured prior to launch (see Sec.~\ref{subsec:SSL_cal}) and the SSL pointing nadir angle $\theta_{\rm SSL}$. With $F_{\rm SSL}$ having been measured at 4.68~m we find $F_{\rm SSL}(d) = (4.68/d)^2 F_{\rm SSL}(4.68~{\rm m})$, with distance $d$ in meters. Then Eq.~\ref{eq:reflectance} becomes
\begin{equation}
r_{\rm C}(\phi) = \frac{(d_1^{-2} - d_2^{-2})^{-1}}{4.68^2} \frac{\pi (I_1 - I_2)}{F_{\rm SSL}(4.68~{\rm m}) \cos\theta_{\rm SSL}}.
\label{eq:final_reflectance}
\end{equation}
The distance to the surface is calculated as $d = h / \cos\theta_{\rm Sp}$ from the altitude $h$ (in meters) and the spectrometer nadir angle ($\theta_{\rm Sp}$) to the center of the footprint on the surface ($\theta_{\rm DLIS} = 21.4^\circ$ and $\theta_{\rm DLVS} = 20.0^\circ$ for the VLNS mode). The DLVS and DLIS wavelength ranges overlap between 800 and 1000~nm. Agreement of the results for both spectrometers in this range would be an important confirmation of the validity of the reconstructed reflectance.

\subsection{Spectrum calibration}
\label{subsec:spec_cal}

The DLVS occupies a section of 200~rows by 20~columns (spectral $\times$ spatial dimension) on the DISR CCD. The Downward Looking Infrared Spectrometer (DLIS) has an array of 150 InGaAs detector elements. Technical details of these instruments together with those of the other DISR hardware can be found in \citet{T02}. The calibration procedure for both spectrometers is described in detail by \citet{T07}. Calibrating the VLNS spectra described in this paper requires attention to some specifics not addressed in the latter paper.

Because DLVS spectra were projected on the CCD slightly warped, with different columns looking at different nadir angles, a geometrical correction is necessary to derive meaningful spectra. Because spectra at both ends of the 20 columns are only partially complete in wavelength, this correction reduces the number of spatial footprints. Hence the 20-column mode, which offers the maximum spatial resolution, yields only 16 complete spectra. Similarly, the (summed) 10-column mode yields 8 spectra, and the 2-column VLNS mode only one. Actually, geometrically correcting 2-column mode spectra involves extrapolation at both ends of the wavelength range. The bias for pre-landing exposures was taken to be constant at 9~DN for each CCD pixel. The post-landing bias was estimated empirically from several exposures chosen to have a wide range in exposure times and a short range in mission times, and was found to be typically below 20~DN. For all spectra considered in this paper the CCD temperature was so low that the dark current was negligible.

In its regular mode of operation the DLIS would open and close the shutter at preset azimuths for a number of rotations, summing the data collected in each of these `regions' on board. In contrast, the VLNS spectra (Table~\ref{tab:dlis_VLNS}) were single exposures with shutter open. Calibrating these involves subtracting a separately recorded dark (shutter closed) spectrum. For this we used DLIS {\bf 208}, acquired at mission time 8860.95~s. The correctly exposed post-landing spectra (Table~\ref{tab:dlis_surface}) were again acquired in the summing mode, and were calibrated as described by \citet{T07}. Merely switching on the SSL imposed a small extra charge on the DLIS array (less than 0.002~W m$^{-2}$ \textmu m$^{-1}$ sr$^{-1}$) for which we corrected.

\subsection{SSL calibration}
\label{subsec:SSL_cal}

\begin{figure}
\centering
\includegraphics[width=6cm,angle=0]{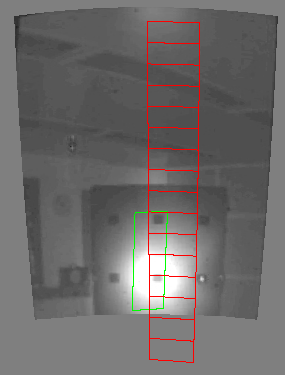}
\caption{A MRI image from the SSL calibration experiments with the DLVS (red) and DLIS (green) slits superposed (polar projection). We see the 95.3$\times$147.3~cm aluminum target with its dark Krylon squares positioned in a room at LPL. The DLVS was operating in the 20-column mode, and the associated 16 footprints are numbered 0 to 15 from top to bottom. The lamp reflection spot is visible in the center of the target. Image brightness and contrast have been adjusted to bring out details in the scene.}
\label{fig:cal_exp_slits}
\end{figure}

An experiment aimed at determining the SSL spectrum was performed at the {\em Lunar and Planetary Laboratory} (LPL) in Tucson, AZ, on 16 August 1996. The DISR\#3 flight model observed a 95.3$\times$147.3~cm (width $\times$ height) painted aluminum target with small (7.6$\times$7.6~cm) dark Krylon squares, positioned at a distance of 4.68 m from the camera and perpendicular to the lamp beam. Measurements were acquired at room temperature by the three cameras and both spectrometers, and housekeeping data were recorded. The SSL current and voltage during the experiment were measured to be nominal, and identical to those during the descent. The DLVS was operating in the 20-column mode, which offers the maximum spatial resolution. Figure~\ref{fig:cal_exp_slits} shows how the instrument viewed its target. The reflective properties of the aluminum and Krylon targets were determined separately with a goniometer at the {\em Laboratoire de Plan\'etologie de Grenoble}, France, in April 2006. Both materials were found to be spectrally flat within a few percent over the full wavelength range of the DLVS and DLIS. It was technically impossible to measure the reflectance at incidence and reflection angle zero, but the reflectance measured with both incidence and reflection angle $20^\circ$ (detector opposite to light source) was found to be a good approximation. The reflectance of the target was taken to be that of the aluminum material, scaled down by a small fraction to account for the presence of the dark Krylon squares. From the intensity $I_{\rm obs}$ observed by the spectrometer and the radiance coefficient $r_{\rm C}^{\rm T}$ of the target we calculate the SSL flux as $F_{\rm SSL}(4.68~{\rm m}) = \pi I_{\rm obs} / r_{\rm C}^{\rm T}(\phi = 0^\circ)$.

\begin{figure}
\centering
\includegraphics[width=8cm,angle=0]{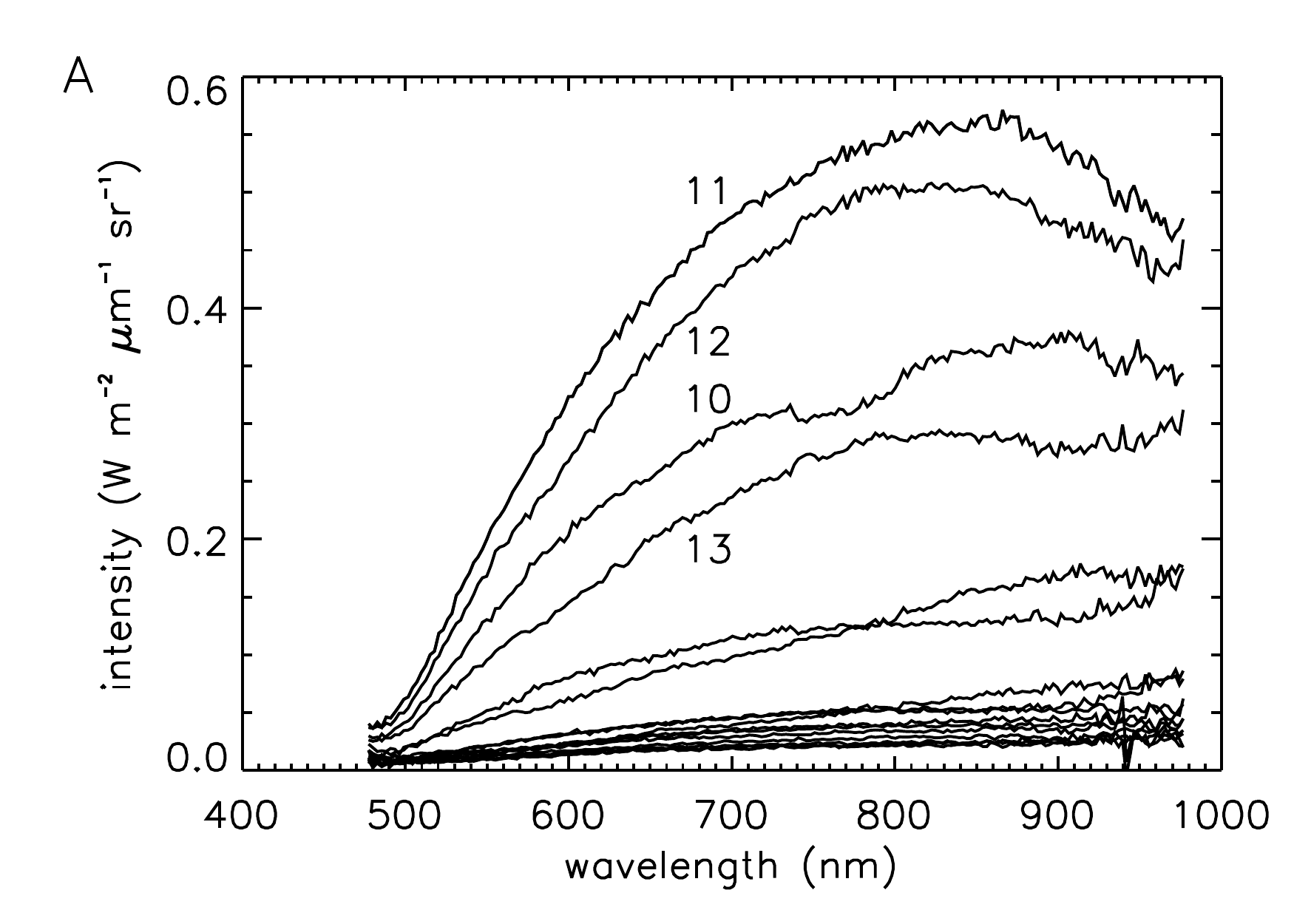}
\includegraphics[width=8cm,angle=0]{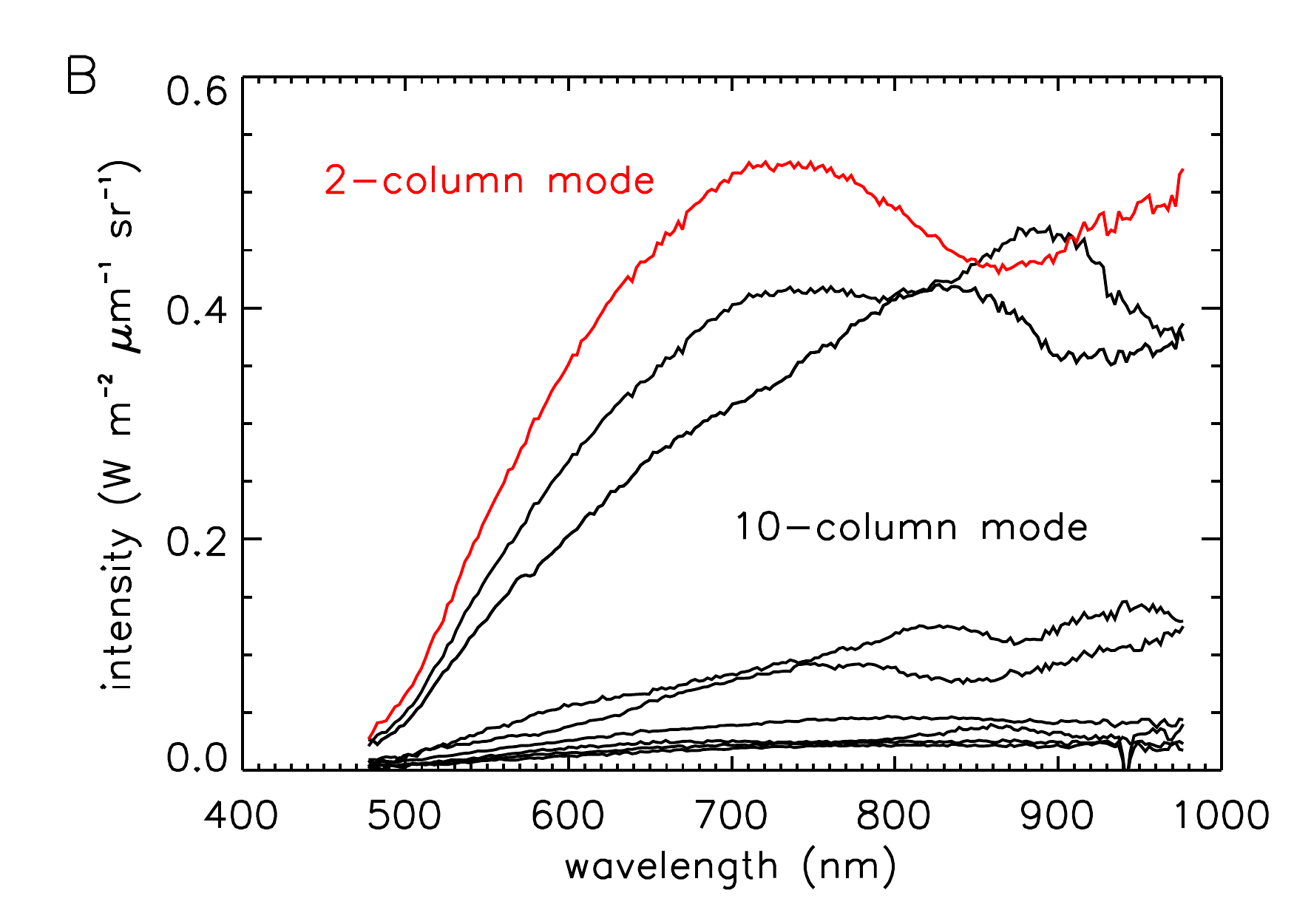}
\includegraphics[width=8cm,angle=0]{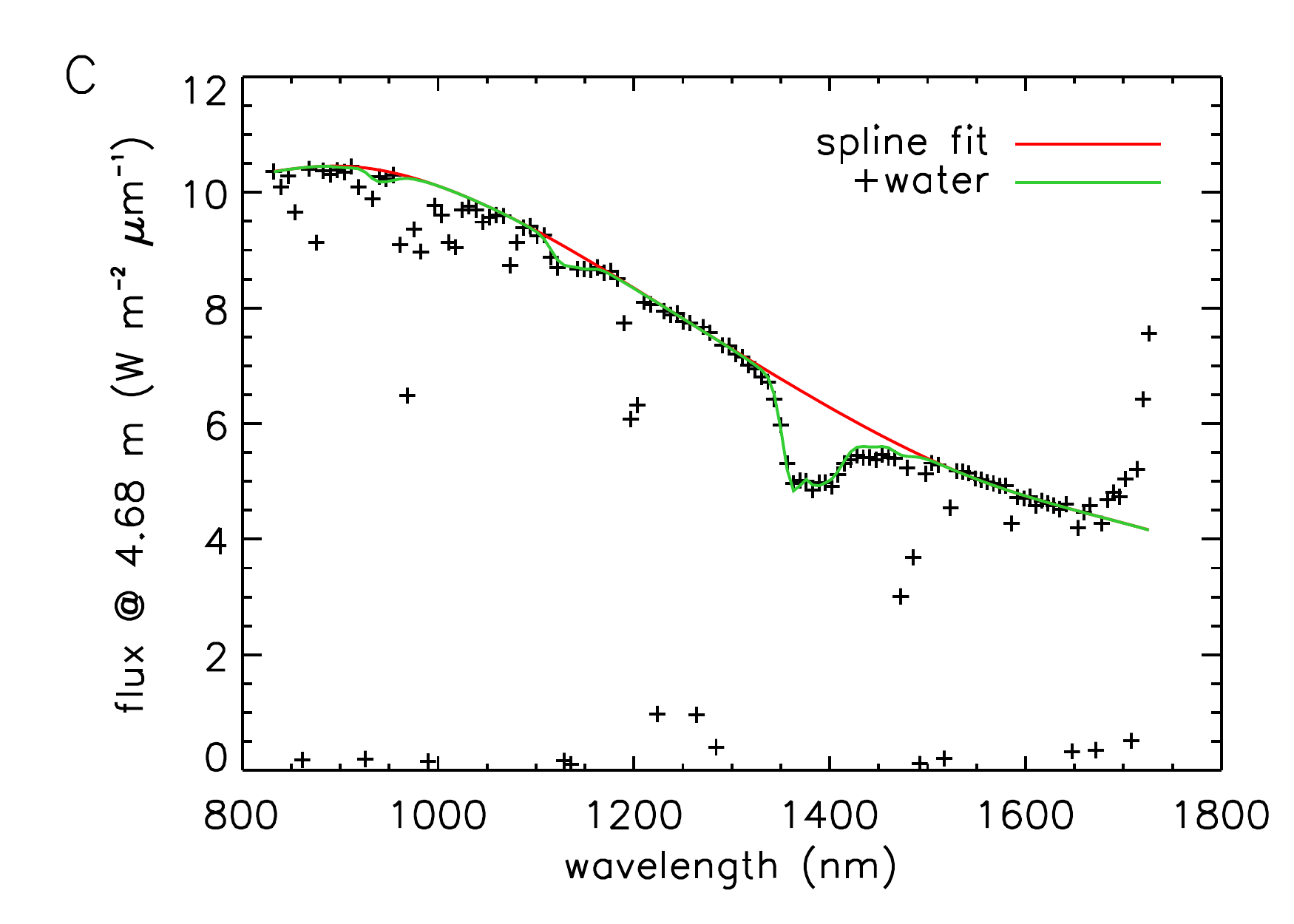}
\includegraphics[width=8cm,angle=0]{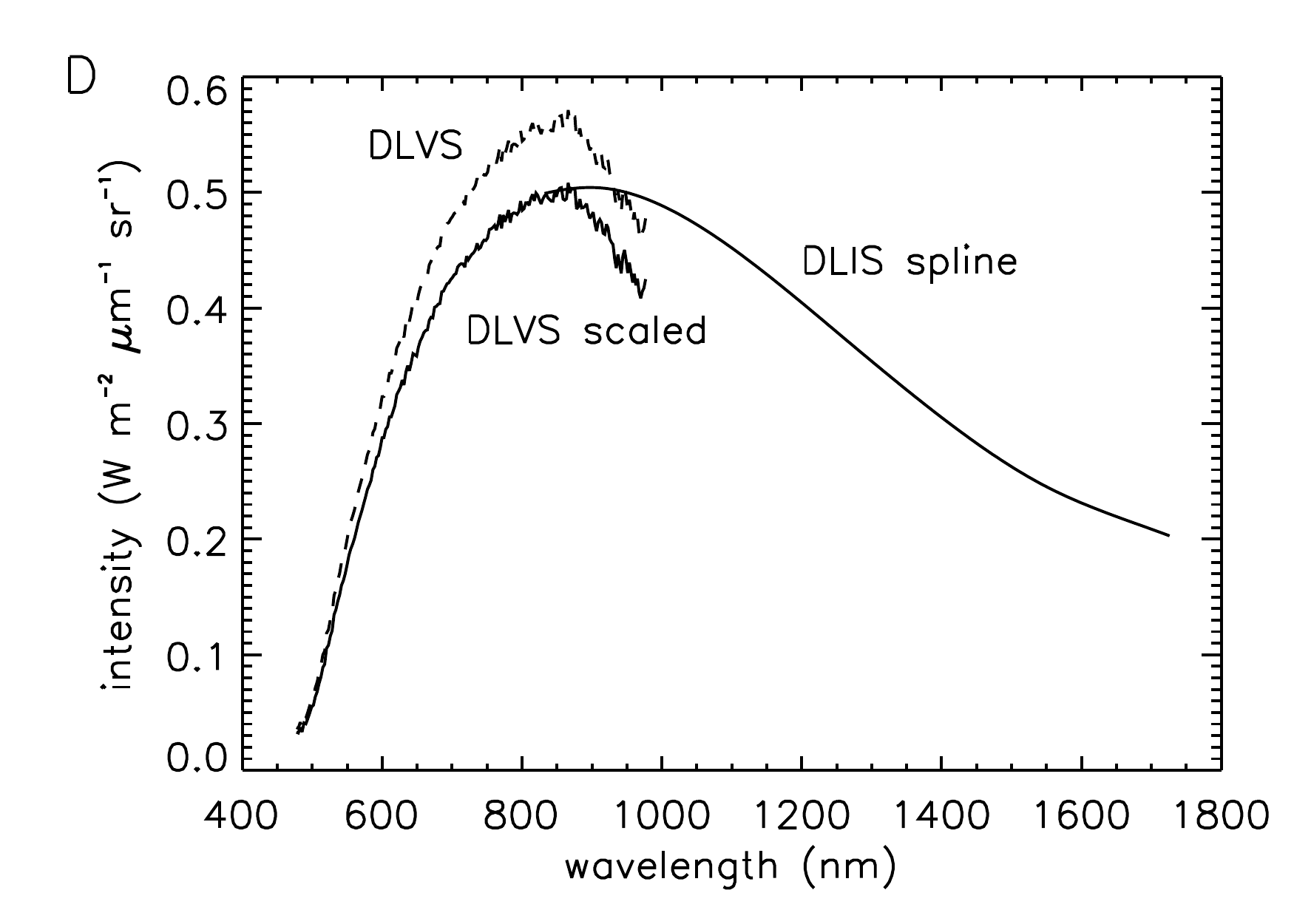}
\includegraphics[width=8cm,angle=0]{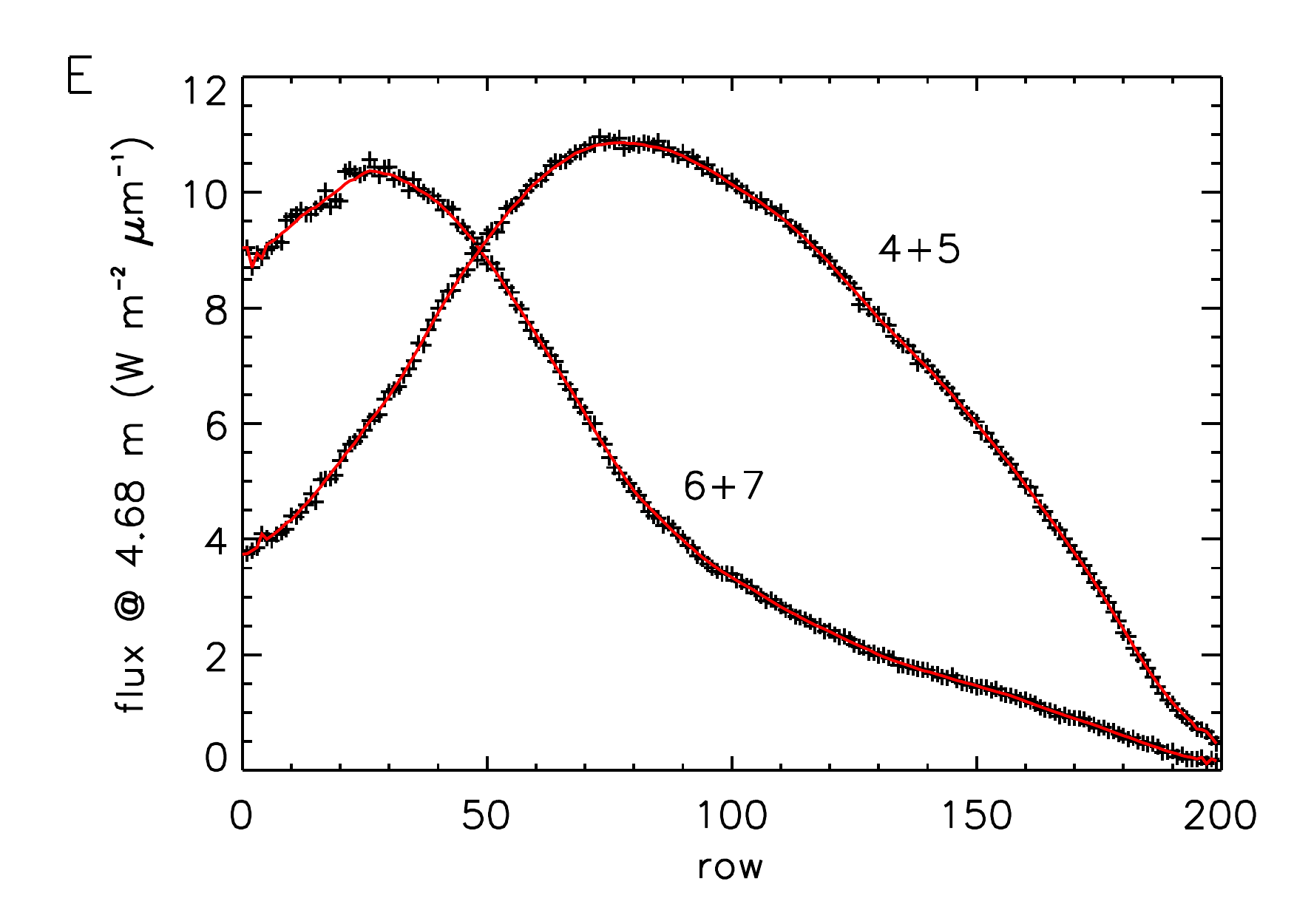}
\caption{The Surface Science Lamp spectrum as observed by the DISR spectrometers in the calibration experiment. {\bf A}: The calibrated spectra of the 16 footprints of the DLVS in the 20-column mode, with some footprint numbers indicated (see Fig.~\ref{fig:cal_exp_slits}). The best reconstruction of the lamp spectrum is that of footprint~11. {\bf B}: The lamp spectrum as observed by the 8~footprints of the DLVS in the 10-column mode. The lamp spectrum as it appears to the DLVS in the 2-column mode in displayed in red. {\bf C}: The adopted SSL spectrum (red) for the DLIS wavelength range was obtained by fitting a spline through the average of two observations (+). The green line includes absorption by atmospheric water vapor (coefficients from the GEISA database). Note the second order response beyond 1600~nm. {\bf D} The full spectrum of the SSL as observed by the DISR spectrometers in the calibration experiment. The dashed line DLVS spectrum is that of footprint~11. The DLIS spectrum is the spline fit from the figure on the left. The drawn line DLVS spectrum has been scaled to match the DLIS spectrum at 850~nm. {\bf E}: The adopted lamp spectrum (red) for the two summed columns (labeled) of the 2-column mode as a function of CCD row number. Wavelength decreases with increasing row number.}
\label{fig:SSL_spectrum}
\end{figure}

The appearance of the SSL spectrum to the DLVS depends strongly on the instrument mode. The 20-column mode, in which the DLVS returns all columns on the CCD, allows for an accurate reconstruction of the true SSL spectrum. The Titan spectra relevant to this paper, however, were acquired in the 2-column mode (pre-landing) and 10-column mode (post-landing), and to reconstruct the reflectance from these spectra we must know how the SSL spectrum appears to the DLVS in these modes. In Fig.~\ref{fig:SSL_spectrum}B we simulate these modes, that is, we adding consecutive columns ourselves and apply the same geometric correction as we do for the Titan spectra. The odd appearance of the lamp spectrum in the 2- and 10-column modes is caused by the finite size of the reflection spot on the CCD; the steep brightness gradients on the spot's edges along the DLVS spatial dimension cannot be reconstructed accurately from only two or ten summed columns. Because the DLIS features only a single footprint, finding the SSL spectrum for the DLIS wavelength range is more straightforward. The observations show the presence of water in the laboratory air, and we fit a cubic spline through the measurements to find the true SSL flux at 4.68~m (Fig.~\ref{fig:SSL_spectrum}C). Combining the DLVS and DLIS observations we arrive at the complete SSL spectrum in Fig.~\ref{fig:SSL_spectrum}D. The DLVS peaks at a higher intensity than the DLIS because a relatively larger fraction of its footprint is covered by the brightest part of the lamp spot (see Fig.~\ref{fig:cal_exp_slits}). If we scale the DLVS spectrum to match the DLIS spectrum, both should agree in the wavelength range where they overlap. Clearly they do not; the DLVS spectrum drops more rapidly beyond 850~nm than the DLIS. The fault probably lies with the DLVS. Either the responsivity, which drops steeply beyond 830~nm, is incorrect, or the geometric correction. The reconstructed lamp spectrum is especially sensitive to the latter because of the aforementioned steep brightness gradient on the CCD. We therefore divide the lamp spectrum out of the observed spectra before the geometric correction, i.e.\ before reducing the two summed columns to a single spectrum. The SSL flux adopted for each of the two (summed) columns is shown in Fig.~\ref{fig:SSL_spectrum}E.

After Huygens' landing the spectrometer footprints and the lamp reflection spot had moved to different locations in the field of view due to the proximity of the surface (compare Figs.~\ref{fig:cal_exp_slits} and \ref{fig:surface_projection}). This led to a completely different brightness distribution on the DLVS section on the CCD. This distribution is unknown because the the flight model did not observe the lamp reflection spot before launch with the target at close range. Since Fig.~\ref{fig:SSL_spectrum}B demonstrates that we cannot accurately reconstruct the SSL spectrum from neither the 2- nor 10-column mode spectra, we do not know how the SSL appeared to the DLVS, and we simply cannot reconstruct the reflectance from post-landing DLVS spectra. Unfortunately, the DLVS was not programmed to switch to the 20-column mode after landing. We attempted to devise a `proximity correction' using the DISR\#2 flight spare camera, with which we observed a target covered by white Krylon paint at far and close range at the LPL in December 2005. First, the target was positioned perpendicular to the SSL beam at 4.46~m, then was moved closer and parallel to the underside of the camera, such that the DLIS window was 46~cm above this surface (as was the situation on Titan, see \citealt{K07}). Spectral changes were observed by the DISR\#2 DLVS, but when we correct the post-landing DISR\#3 reflectance, the resulting spectra look completely different from the pre-landing result. Hence we deem the correction unreliable, being too sensitive to the summing and interpolation involved in the DLVS calibration. We do not expect the DLIS to observe changes in the reflected lamp spectrum after landing. Nevertheless, we noticed minor changes in the lamp spectrum observed by the DISR\#2 DLIS after moving to the target at close range; the most significant one being a drop in intensity from 1350~nm onward, peaking at 5\% at 1500~nm. In addition, assuming the lamp flux scales from 4.68 to 4.46~m according to 1/$d^2$, we find that at close range the SSL was perceived to be 3.04 times weaker as would be expected from the calibration experiment (again, scaling according to 1/$d^2$); a clear consequence of parallax.

\section{Results}
\label{sec:results}

\subsection{DLVS}
\label{sec:DLVS_results}

\begin{figure} \centering
\includegraphics[width=8cm,angle=0]{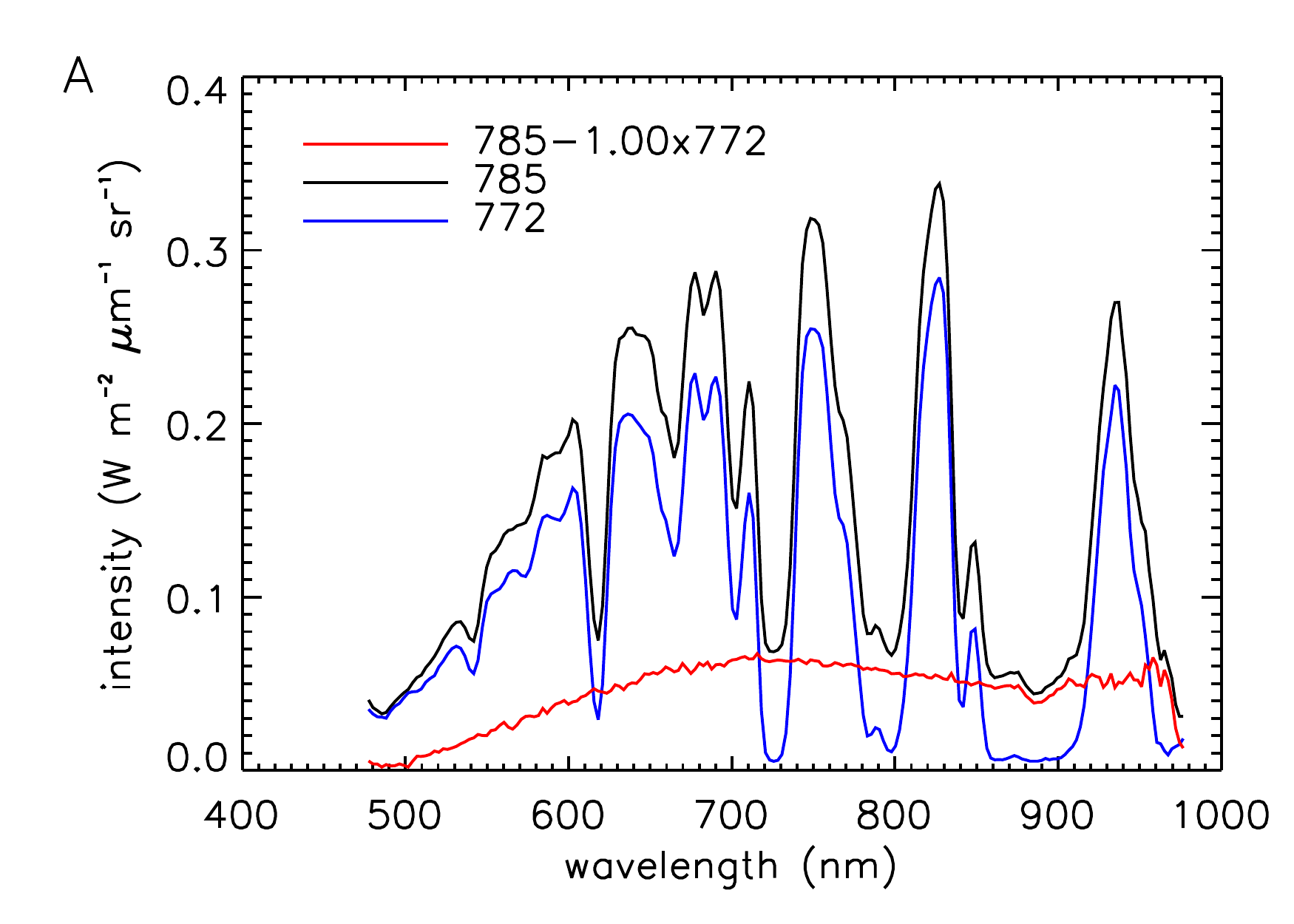}
\includegraphics[width=8cm,angle=0]{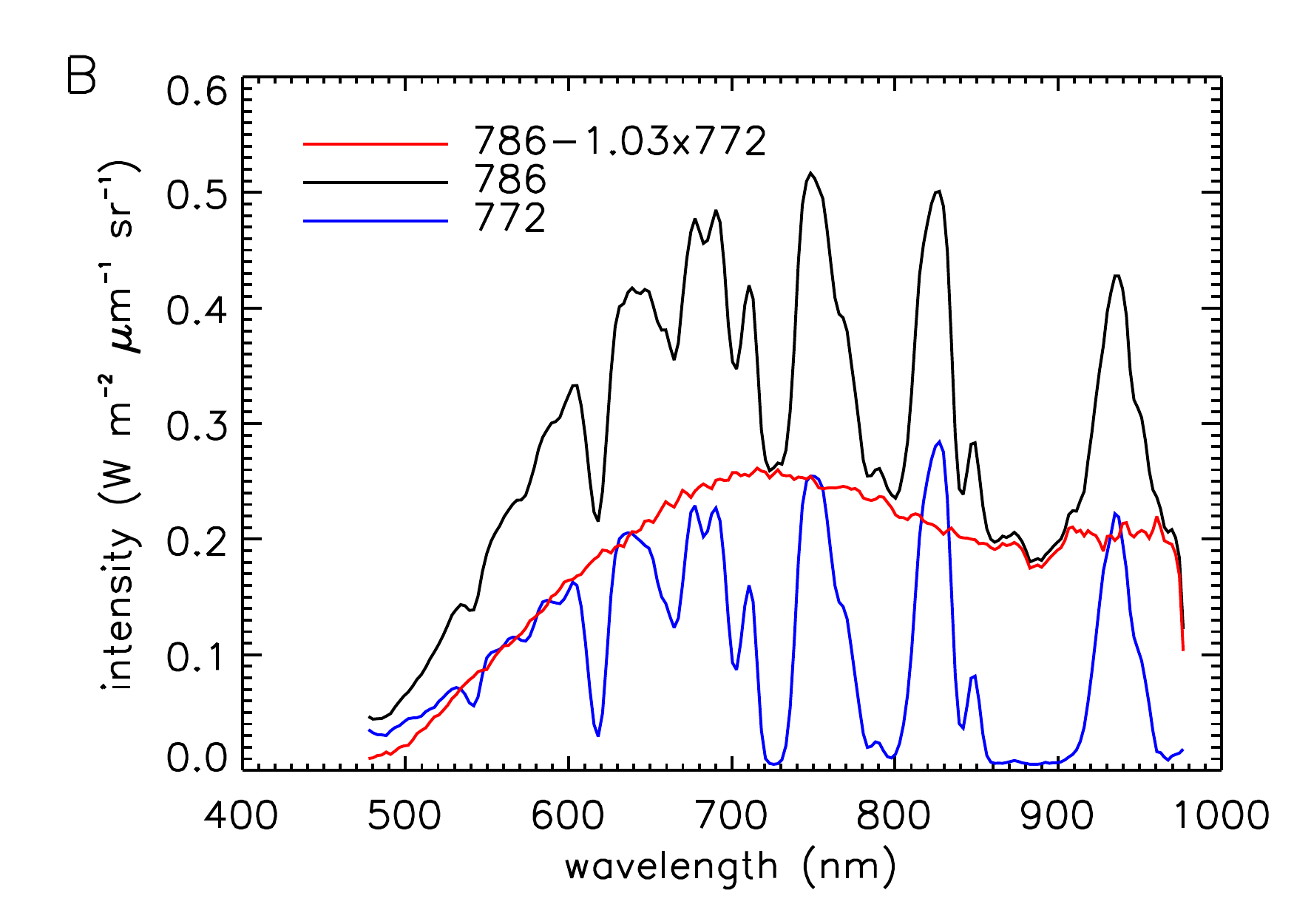}
\caption{The procedure to reconstruct the reflectance is illustrated for the two pre-landing DLVS spectra in which a lamp signal is clearly present ({\bf A}: {\bf 785}, {\bf B}: {\bf 786}). The lamp-only spectrum (red) is constructed by subtracting a background spectrum (blue), multiplied by a constant, from the observed spectrum (black). Compare the lamp-only spectra with the simulated 2-column mode lamp spectrum in Fig.~\ref{fig:SSL_spectrum}B. Note that we do not divide the lamp-only spectrum {\em as shown} by the lamp spectrum to find the reflectance, but divide out the lamp {\em before} converting the two summed columns of the lamp-only spectrum to a single spectrum by means of the geometric correction.}
\label{fig:DLVS_difference_spectra}
\end{figure}

Two pre-landing DLVS spectra which show clear evidence for lamp light are {\bf 785} and {\bf 786}. Finding a suitable background spectrum for the reflectance reconstruction is not straightforward. Spectrum {\bf 772} appears to be a good choice as it was recorded at approximately the same solar phase angle as {\bf 785} and {\bf 786} (see Table~\ref{tab:dlvs_VLNS}). But its altitude is more than twice as high as, for example, {\bf 779}, and the absorption by haze particles in the last hundred meters may not be negligible. Since this predominantly affects the blue end of the spectrum, {\bf 779} may be the better choice at these wavelengths. Neither spectrum is perfect, and we try out both. Figure~\ref{fig:DLVS_difference_spectra} shows the spectra involved in the reconstruction.

\begin{figure}
\centering
\includegraphics[width=8cm,angle=0]{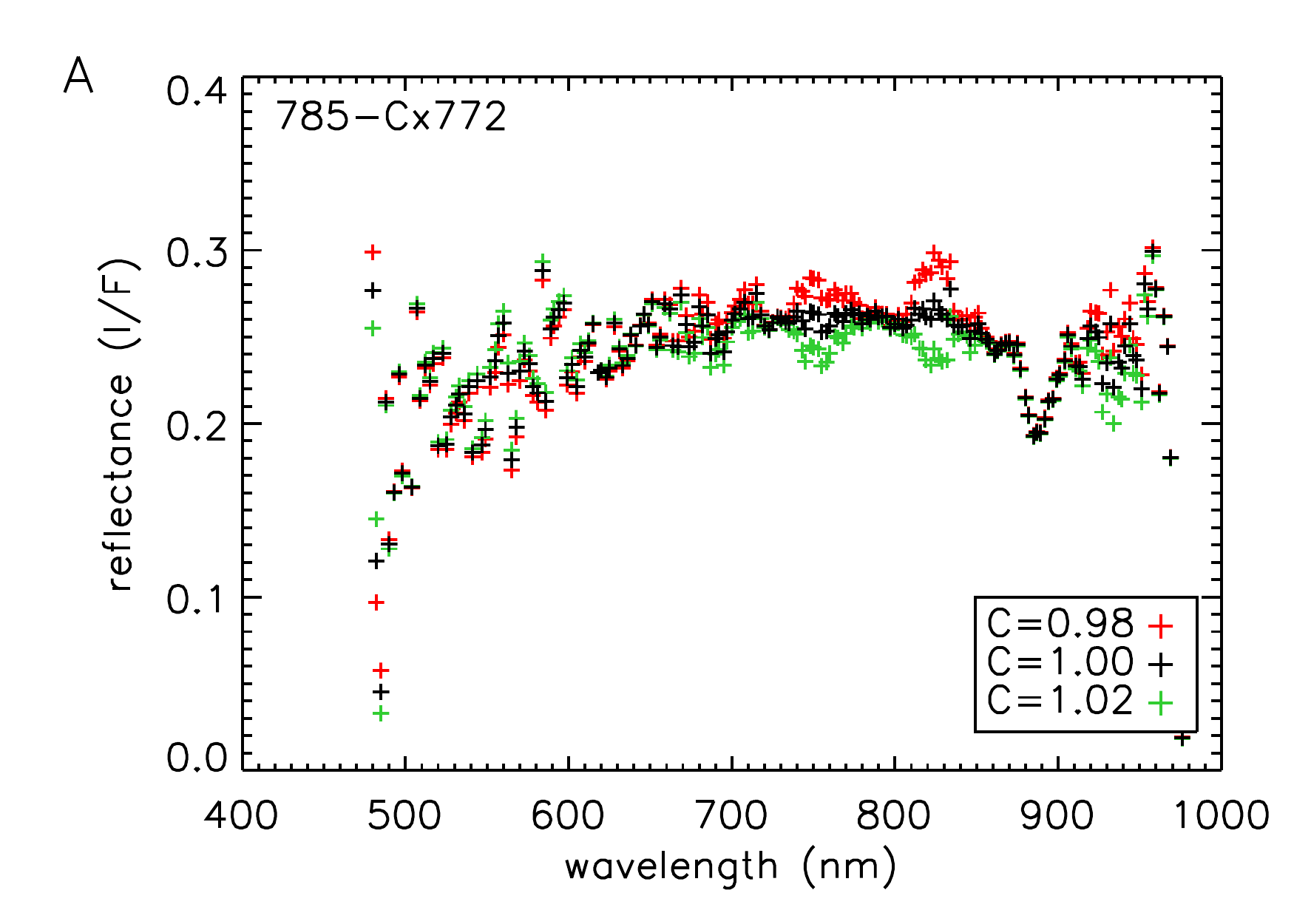}
\includegraphics[width=8cm,angle=0]{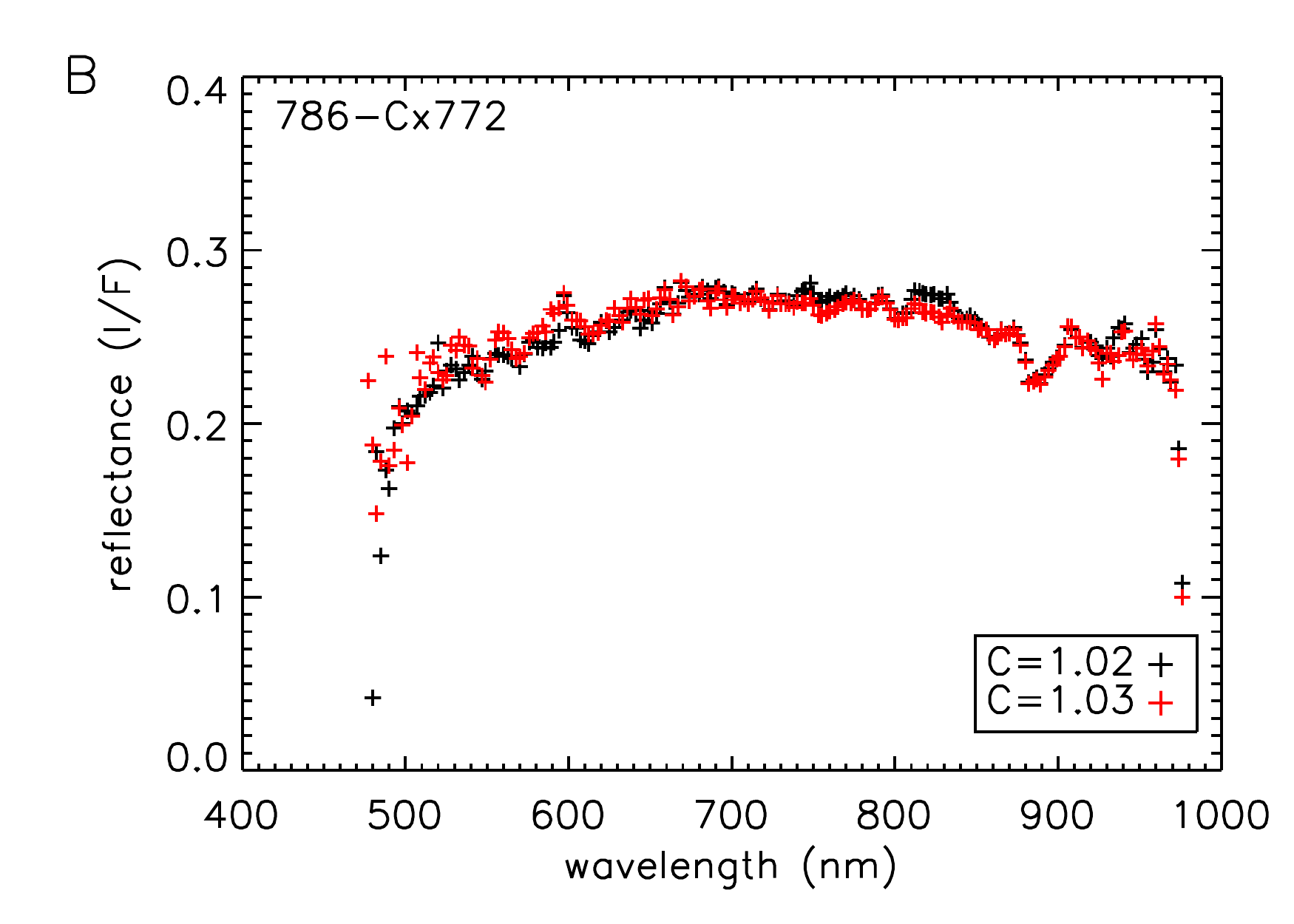}
\caption{The visual surface reflectances derived from the one-before-last ({\bf A}: {\bf 785}) and last ({\bf B}: {\bf 786}) DLVS spectra before landing. The figures show how varying the background spectrum by a few percent affects the reconstruction; the legend lists the factor by which the background was multiplied before subtraction.}
\label{fig:reflectance_785_786}
\end{figure}

It turns out that the choice of background spectrum ({\bf 772} or {\bf 779}) hardly affects the end result, so we use {\bf 772} because of the better phase angle. What matters more is how much background we subtract. Figure~\ref{fig:reflectance_785_786}A shows how changing the background by a few percent affects the reflectance of {\bf 785} in the methane windows; subtracting too much background creates ``absorption lines'' in the methane windows, whereas the opposite creates ``emission lines''. Hence our strategy is varying the background to achieve maximum smoothness of the reflectance. The effect on spectrum {\bf 786} is more subtle (Figure~\ref{fig:reflectance_785_786}B); subtracting too much background makes the red slope below 700~nm less smooth, too little creates spurious emission lines. Overall, the reflectances reconstructed from {\bf 786} and {\bf 786} are similar; an increase towards 800~nm and a decrease beyond. Naturally, the {\bf 785} reflectance is noisier, but the only significant difference is the depth of the 890~nm methane absorption band, as expected since {\bf 785} has an optical path length about four times longer.

\begin{figure}
\centering
\includegraphics[width=8cm,angle=0]{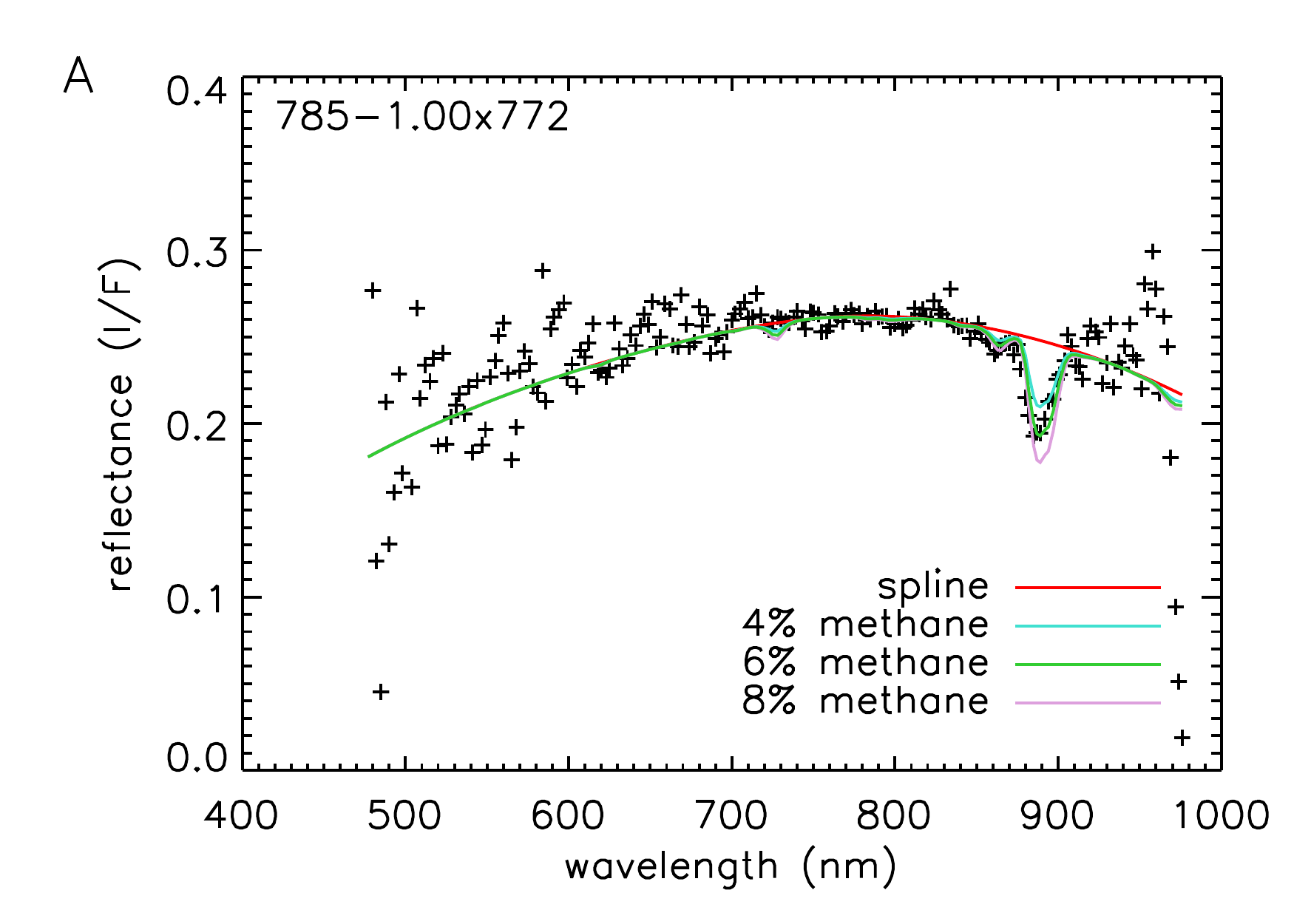}
\includegraphics[width=8cm,angle=0]{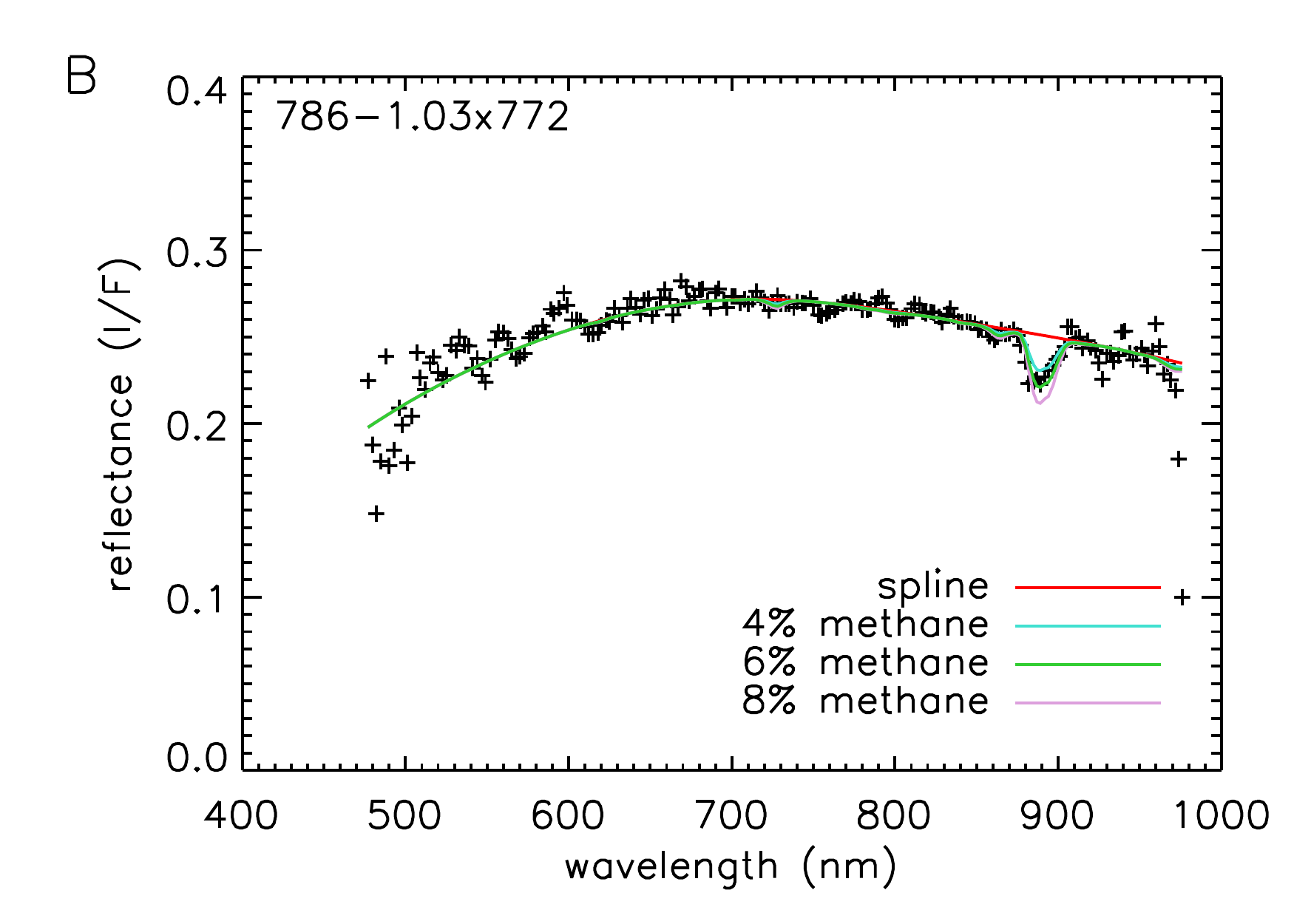}
\caption{The reflectance reconstructed from the pre-landing DLVS spectra can be modeled by superposing methane absorption on a spline fit (red line). Reasonable fits are achieved for a $6\pm2$\% methane mixing ratio. {\bf A}: {\bf 785}, {\bf B}: {\bf 786}.}
\label{fig:vis_methane_fit}
\end{figure}

How reliable are these reconstructions? In Sec.~\ref{subsec:SSL_cal} we found that there may be an error associated with the geometric correction of the lamp spectrum at the high wavelength end. This should not affect our reconstruction, since we divide out the lamp spectrum before the two summed columns are converted to a single spectrum. If the surface is uniform and lamp light has been divided out correctly, both these columns should be identical. However, the reflectance associated with summed columns 6+7 of both {\bf 785} and {\bf 786} is typically 25-30\% higher than that of summed columns 4+5. But as {\bf 786} was acquired at a distance to the surface closest to that of the target distance in the calibration experiments, this is cause for concern. It may indicate that we have not correctly accounted for the lamp signal. We model the optimal reflectance reconstructions by superposing methane absorption on a spline fit representing the true surface reflectance. Figure~\ref{fig:vis_methane_fit} shows that the 890~nm methane band in both pre-landing spectra can be modeled reasonably well with a methane mixing ratio of $6\pm2$\%. Apart from this methane band, the spectra appear to show additional absorption lines, for example around 800~nm. However, these do not appear in both summed columns, and must be considered artifacts.

\begin{figure} \centering
\includegraphics[width=8cm,angle=0]{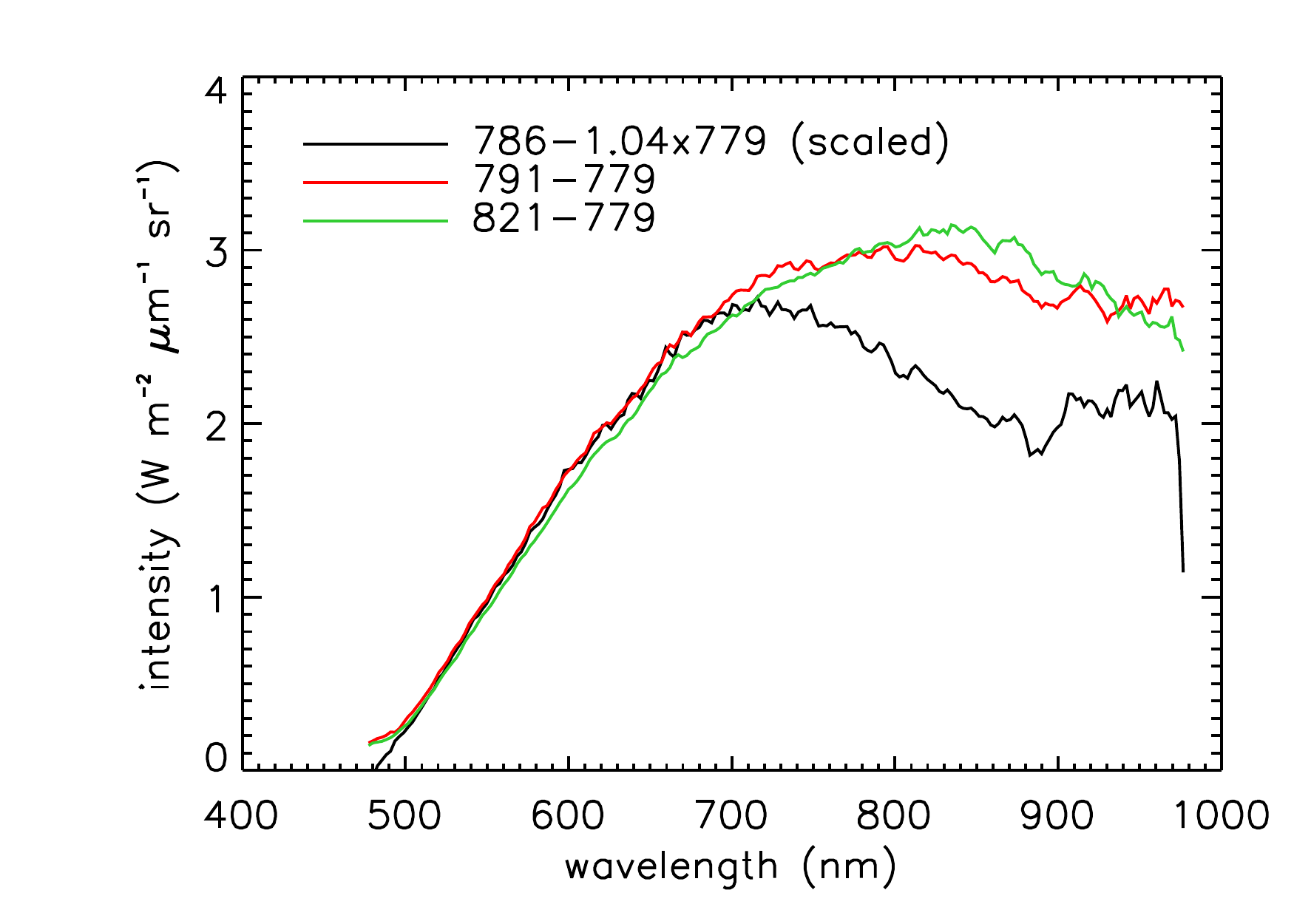}
\caption{The reflectance cannot be reconstructed reliably from post-landing spectra due to the unknown shape of the lamp spectrum. The figure compares post-landing spectra {\bf 791} (a 2-column mode spectrum) and {\bf 821} (footprint 6 of a 10-column mode spectrum) with pre-landing spectrum {\bf 786}, after background subtraction. Even though in principle the footprints of all spectra cover the same type of terrain, the spectra appear different due to the different modes of acquisition.} \label{fig:reflectance_791}
\end{figure}

It proved not possible to reconstruct the reflectance from post-landing spectra. With the surface so close the DLVS perceived the lamp spectrum completely different compared to the situation before landing (Fig.~\ref{fig:reflectance_791}). We deem the accurate reconstruction of the shape of the SSL spectrum and the reflectance impossible, especially because of the on-board summing of the CCD columns. The relevance of the post-landing spectra is therefore limited. The examples shown in Fig.~\ref{fig:reflectance_791} merely demonstrate the absence of absorption lines.

\subsection{DLIS}
\label{DLIS}

\begin{figure}
\centering
\includegraphics[width=8cm,angle=0]{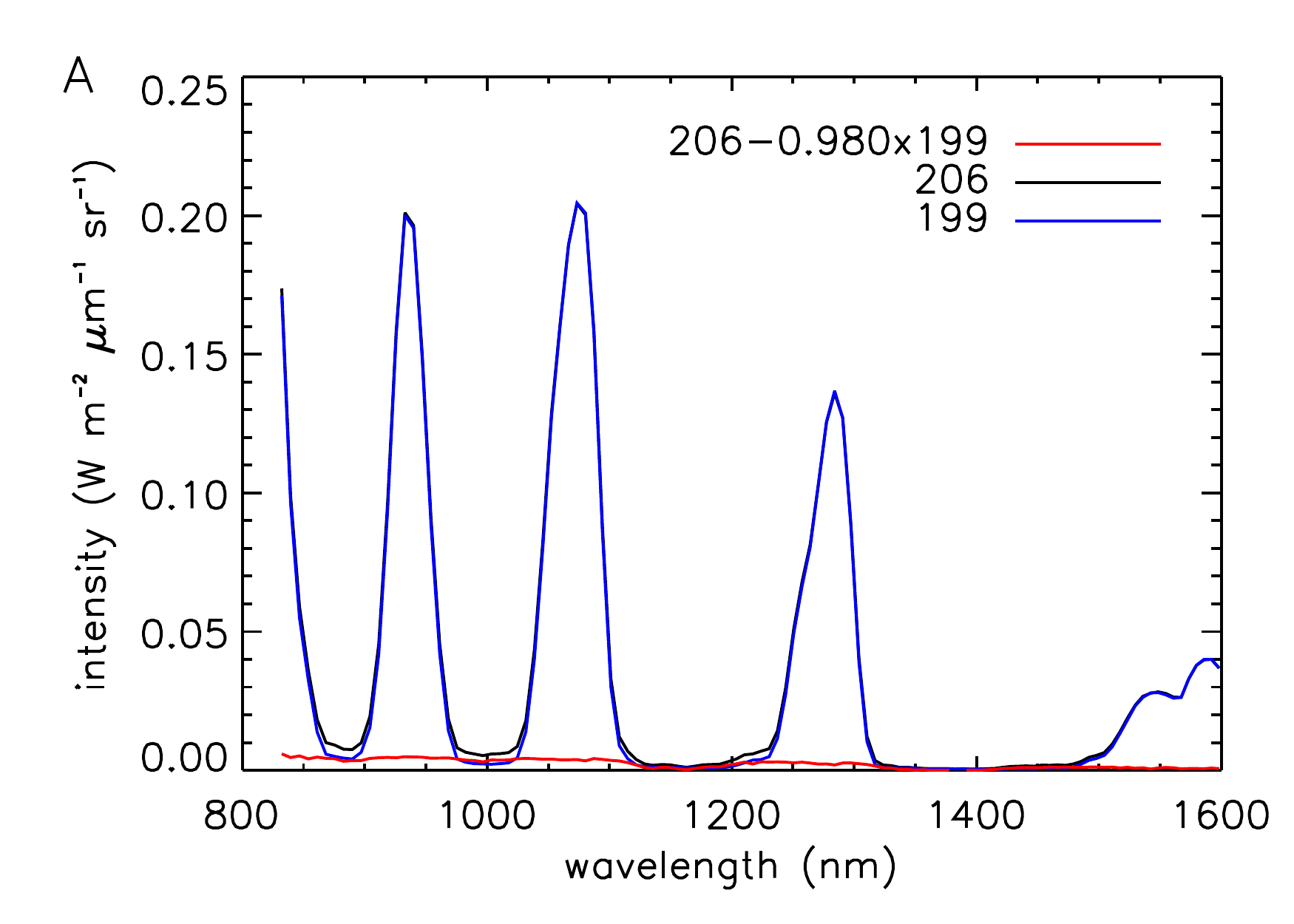}
\includegraphics[width=8cm,angle=0]{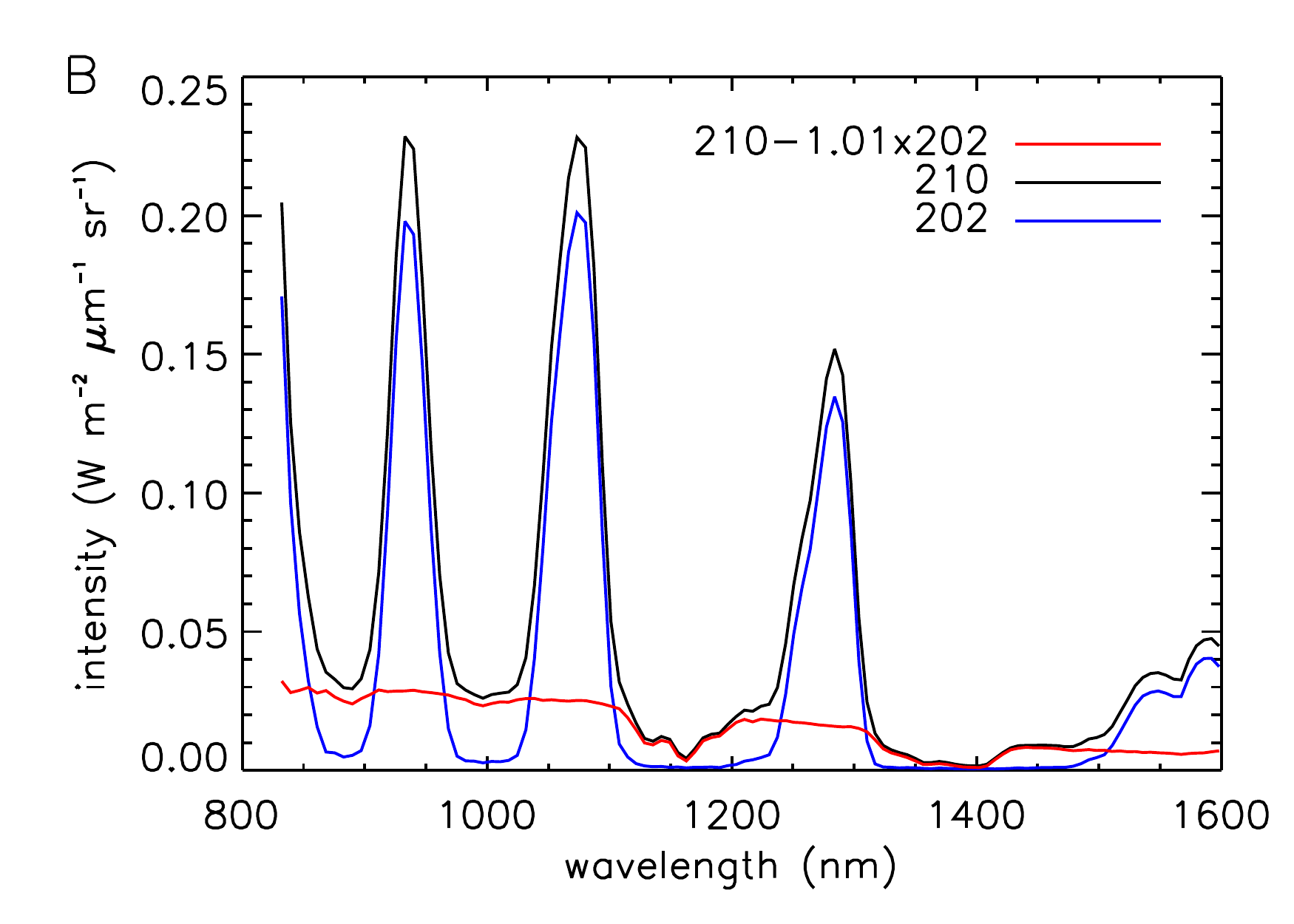}
\caption{The procedure to reconstruct the reflectance is illustrated for the two pre-landing DLIS spectra in which a lamp signal is clearly present ({\bf A}: {\bf 206}, {\bf B}: {\bf 210}). The lamp-only spectrum (shown in red) is constructed by subtracting the background spectrum (blue), multiplied by a constant, from the observed spectrum (black). The surface reflectance is obtained by dividing the lamp-only spectrum by the lamp spectrum in Fig.~\ref{fig:SSL_spectrum}C.} \label{fig:DLIS_difference_spectra}
\end{figure}

Two pre-landing DLIS spectra which show clear evidence for lamp light are {\bf 206} and {\bf 210}. Again we are faced with the task of finding a suitable background spectrum for the reflectance reconstruction. Because it was recorded at approximately the same solar phase angle as {\bf 206} and {\bf 210} (see Table~\ref{tab:dlis_VLNS}), spectrum {\bf 202} appears to be the best candidate. We also try out an earlier spectrum, {\bf 199}, which was recorded at a slightly smaller phase angle. Figure~\ref{fig:DLIS_difference_spectra} shows all spectra involved in the reconstruction.

\begin{figure}
\centering
\includegraphics[width=8cm,angle=0]{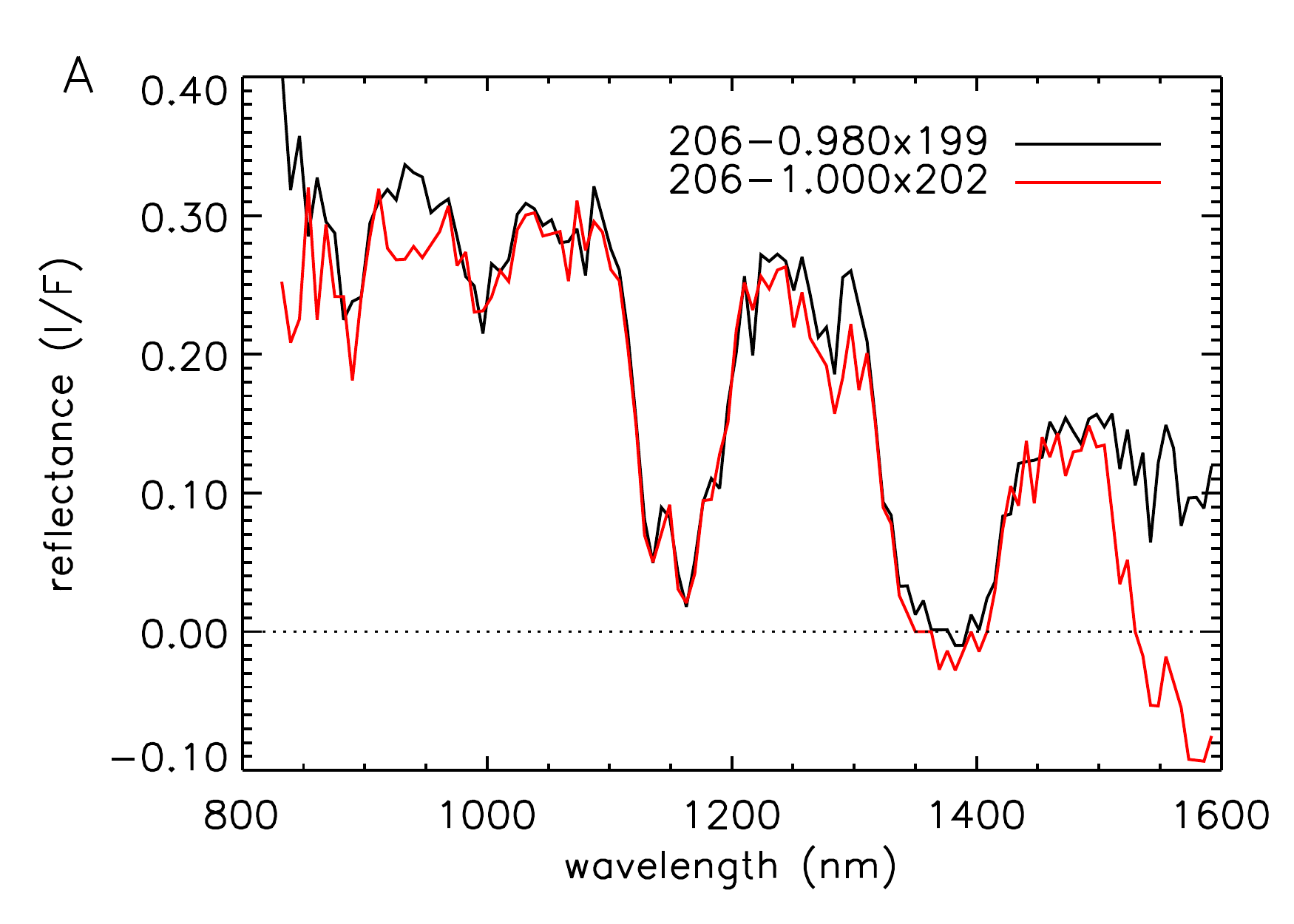}
\includegraphics[width=8cm,angle=0]{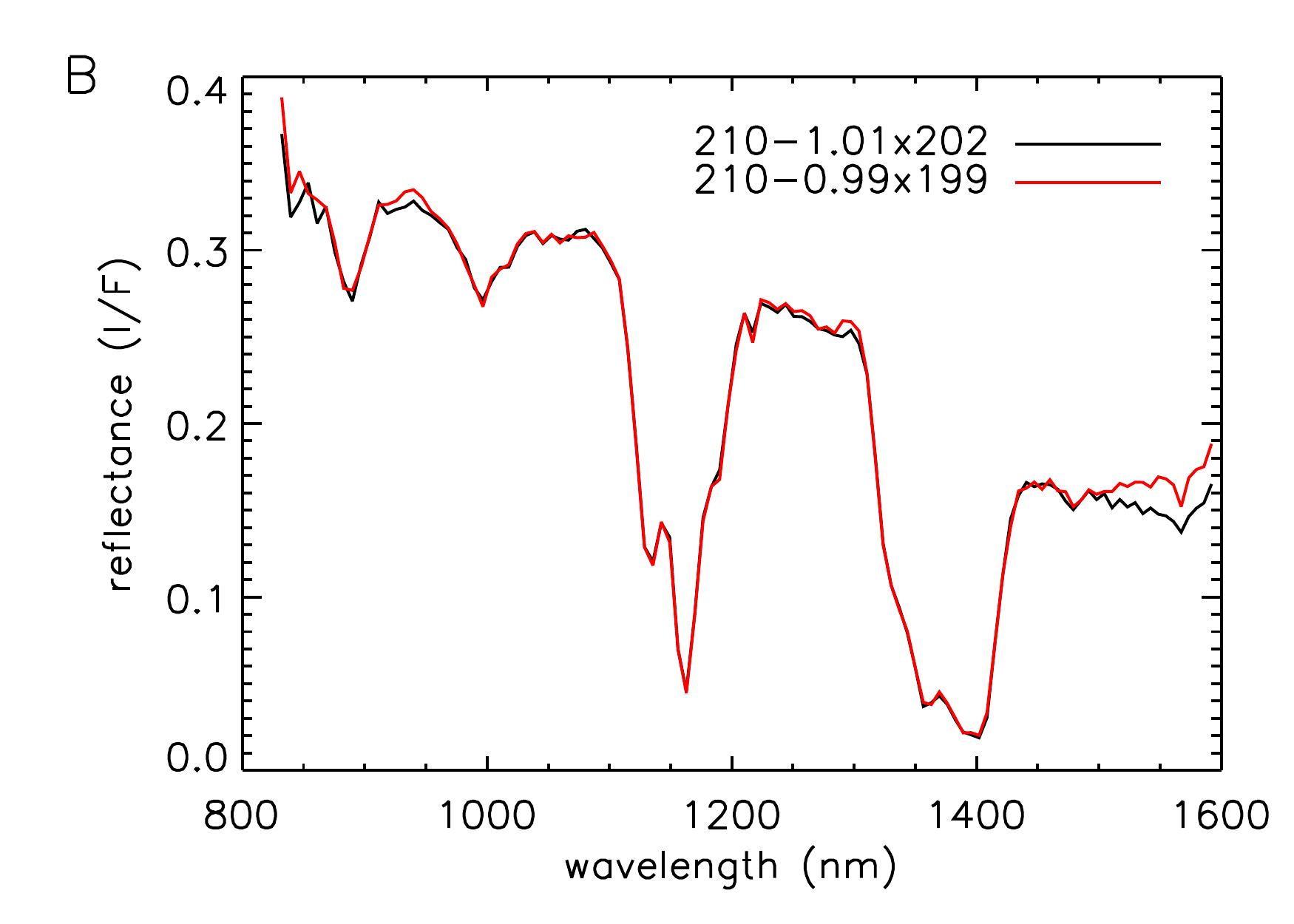}
\includegraphics[width=8cm,angle=0]{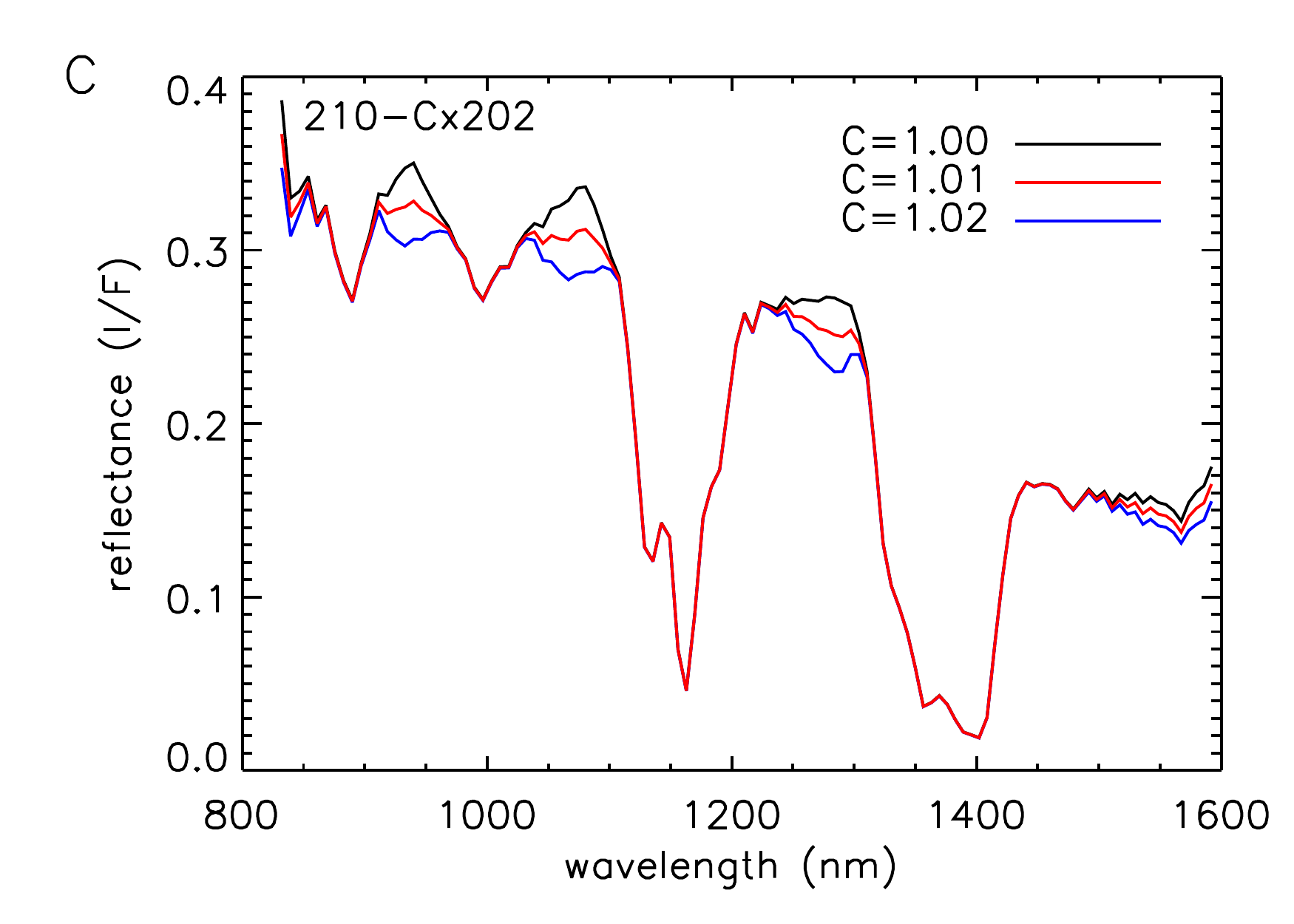}
\caption{The reflectance reconstructions derived from the one-before-last ({\bf A}: {\bf 206}) and last ({\bf B}: {\bf 210}) DLIS spectra before landing using different background spectra. In {\bf C} background spectrum {\bf 202} was varied by a few percent to show how this affects the reflectance in the methane windows. The legend lists the factors by which the background was multiplied before subtraction from {\bf 210}.}
\label{fig:reflectance_206_210}
\end{figure}

\begin{figure}
\centering
\includegraphics[width=8cm,angle=0]{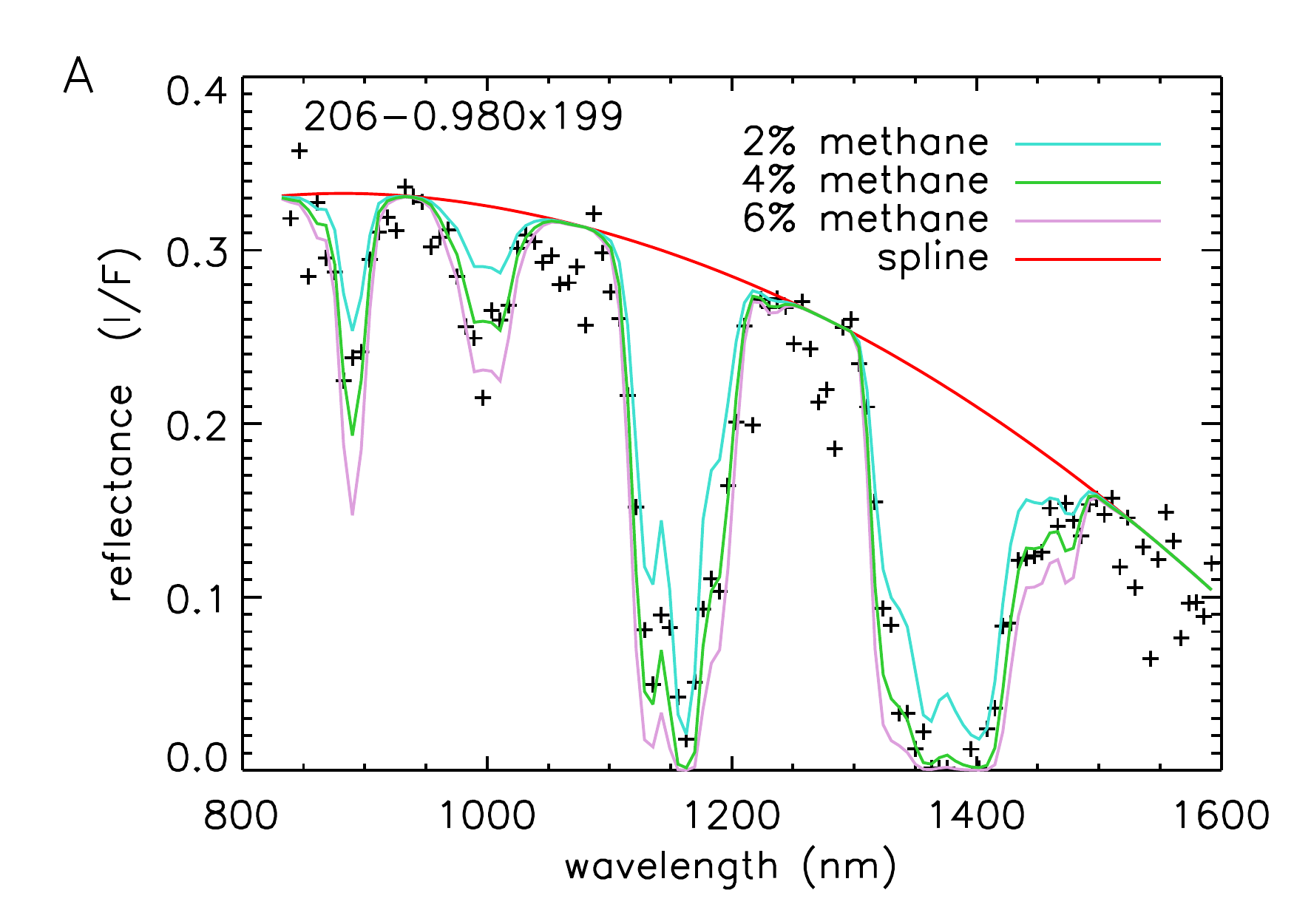}
\includegraphics[width=8cm,angle=0]{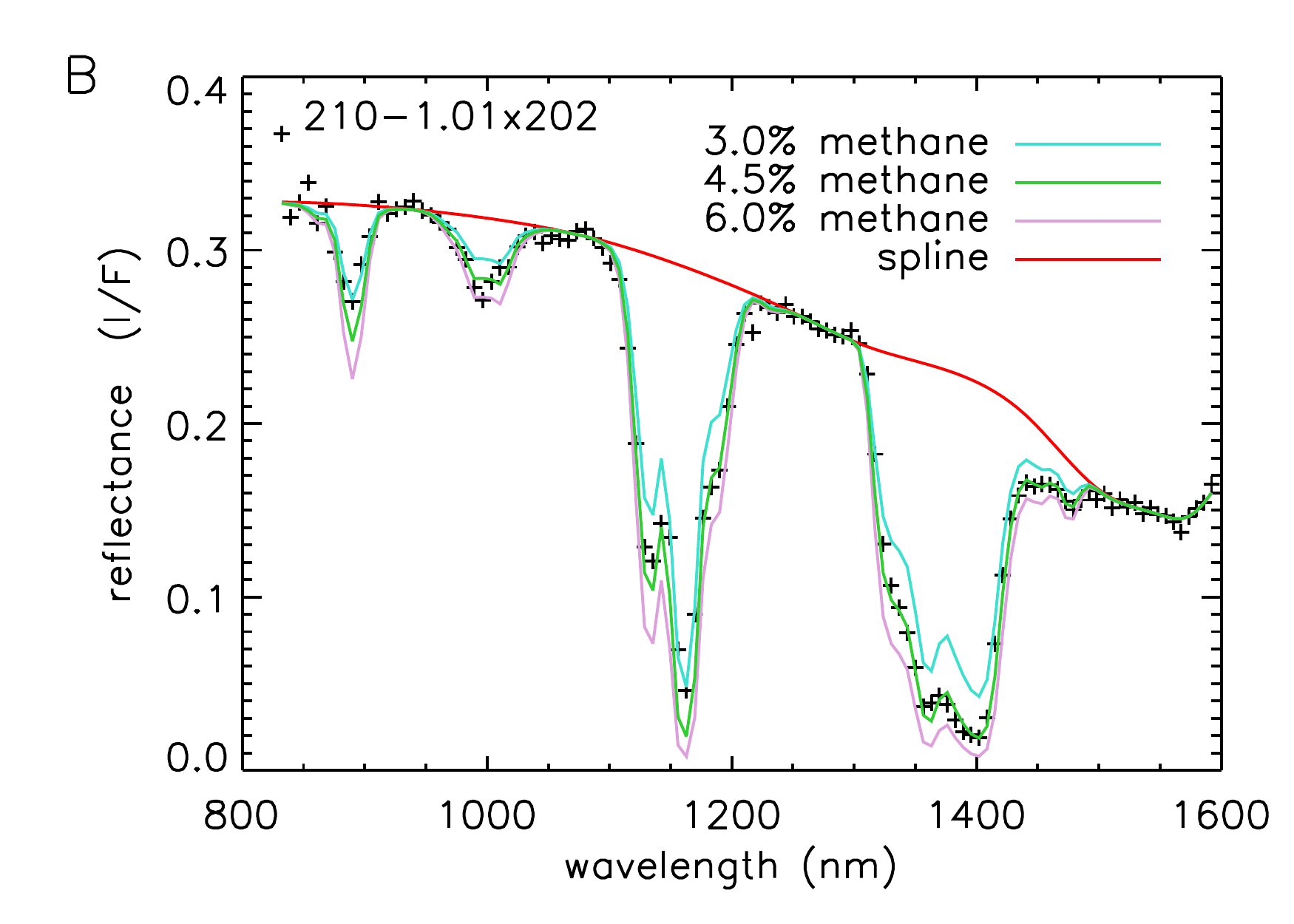}
\caption{The reflectance reconstructed from the pre-landing DLIS spectra can be modeled by superposing methane absorption on a spline fit through the methane windows (red line). {\bf A}: The reflectance reconstructed from {\bf 206}, albeit noisy, is best modeled with a 4\% methane mixing ratio. {\bf B}: The reflectance reconstructed from {\bf 210} is modeled well with a 4.5\% methane mixing ratio. The slope around 1450~nm was introduced into the spline to achieve a good fit.}
\label{fig:IR_methane_fit}
\end{figure}

The sensitivity of the reconstruction to the choice of background is illustrated in Fig.~\ref{fig:reflectance_206_210}. Like for the DLVS we find optimally smooth reflectances by varying the background spectra a few percent (Fig.~\ref{fig:reflectance_206_210}C): subtracting too much creates absorption bands in the methane windows, whereas too little leads to methane bands with an unphysical shape. The reflectance is similar for all reconstructions, decreasing steadily from 800 to 1500~nm. Methane absorption bands due to the intervening atmosphere are apparent, with a saturated 1400~nm band for {\bf 206}. Only the reflectance beyond 1500~nm appears to be sensitive to the choice of background spectrum (it can even become negative for {\bf 206}), and is therefore not fully reliable. We model the optimal reflectance reconstructions by superposing methane absorption on a spline fit through the methane windows representing the true surface reflectance. As Fig.~\ref{fig:IR_methane_fit} shows, the {\bf 206} and {\bf 210} reflectances are best modeled with a $4\pm1$\% and $4.5\pm0.5$\% methane mixing ratio, respectively. Spectrum {\bf 206} is quite noisy, and a smooth spline through the methane windows suffices as a model for the surface reflectance. In the high S/N spectrum {\bf 210} we need to introduce a slope around 1450~nm into the spline model to achieve a satisfactory fit. The resulting feature at 1500~nm likely represents an absorption band. Note that the depression is not required when we use the \citet{S93} methane absorption coefficients instead of those from \citet{I06}.

\begin{figure}
\centering
\includegraphics[width=8cm,angle=0]{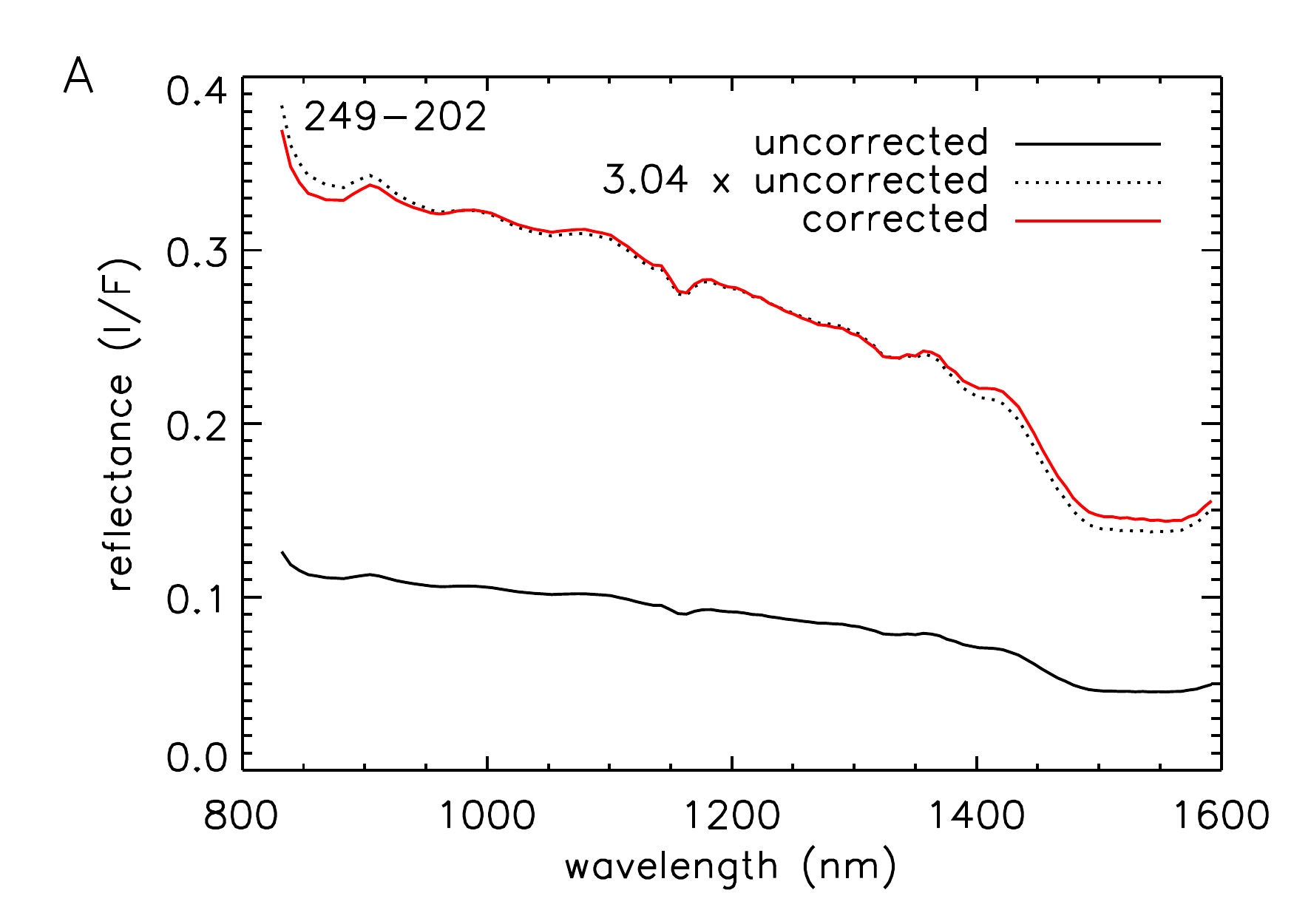}
\includegraphics[width=8cm,angle=0]{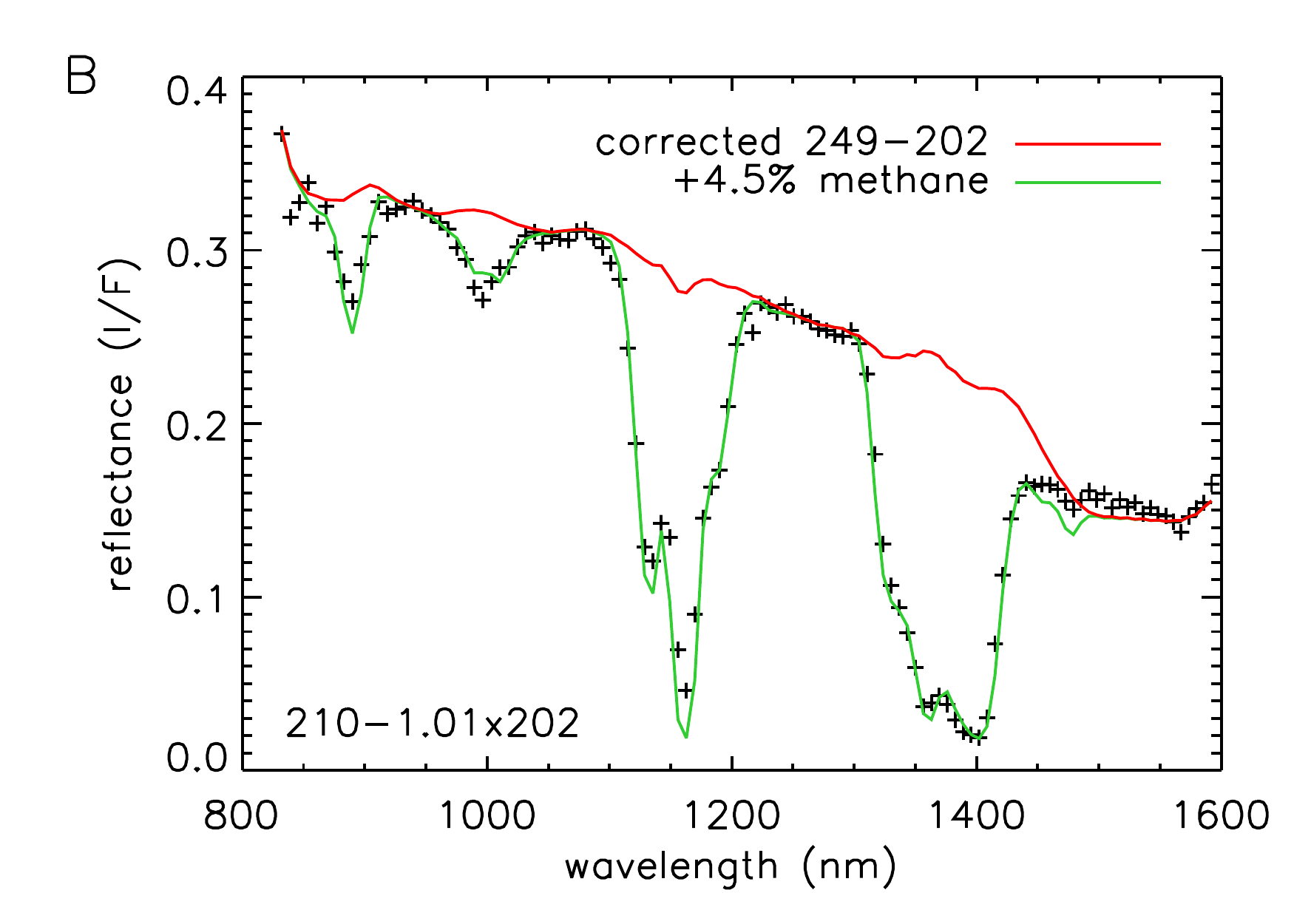}
\includegraphics[width=8cm,angle=0]{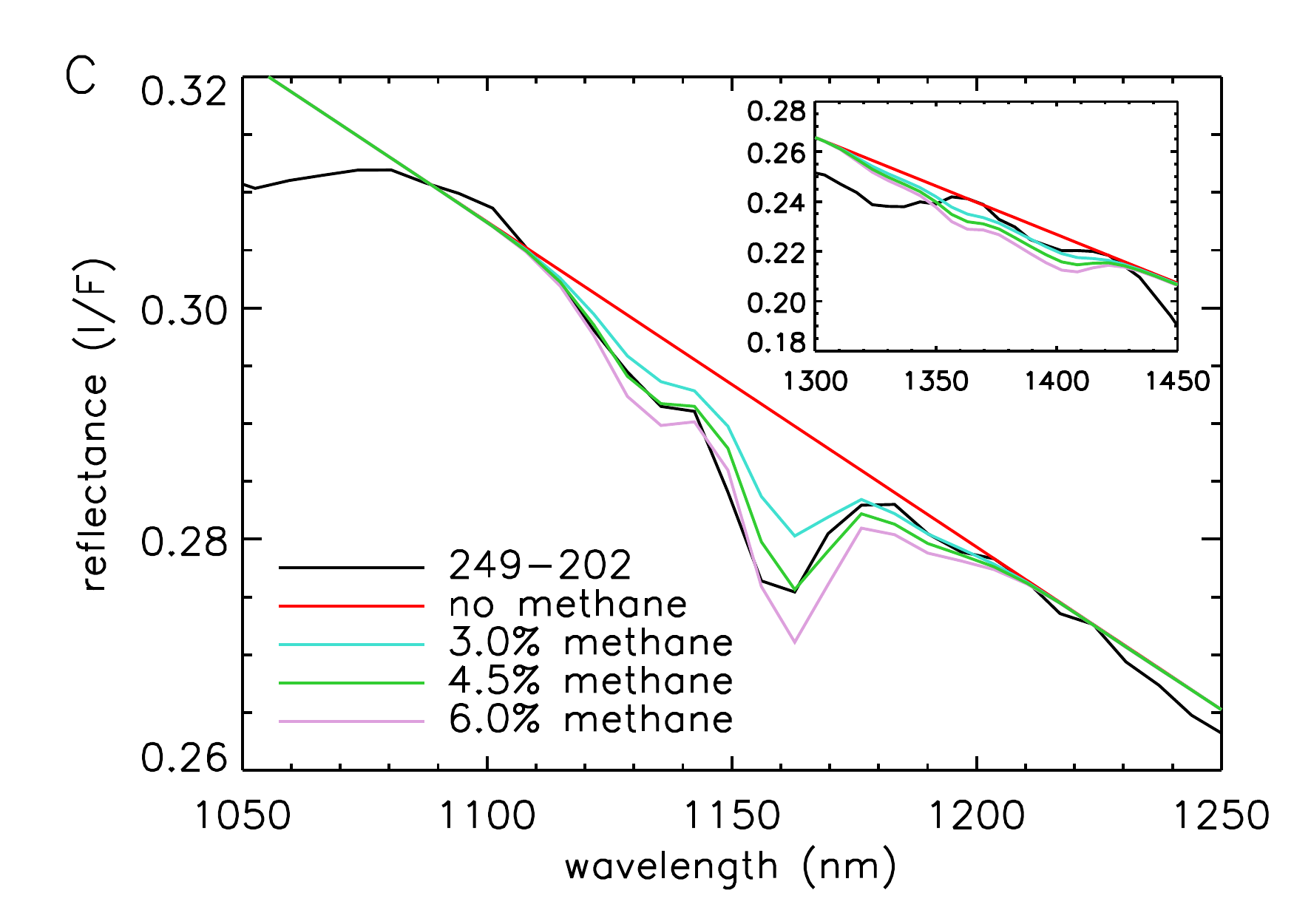}
\caption{The reflectance reconstructed from post-landing DLIS spectrum {\bf 249}. {\bf A}: Correcting the post-landing reflectance for the proximity of the surface using results from the flight spare (DISR\#2) experiment makes it increase by a factor of three, and changes the shape of the red end slightly. {\bf B}: The corrected post-landing reflectance from {\bf A} agrees very well with pre-landing spectrum {\bf 210}, except around 1450~nm. The green line adds 4.5\% atmospheric methane absorption to reflectance {\bf 249}. {\bf C}: The 1160~nm methane absorption band in the reflectance spectrum is modeled well with a 4.5\% methane mixing ratio, assuming the DLIS is located 46 cm above the surface. The continuum is defined linear. The situation is unclear for the 1400~nm absorption complex ({\bf inset}).} \label{fig:reflectance_249}
\end{figure}

After landing the DLIS continued to operate, peering straight into the lamp reflection spot (see Fig.~\ref{fig:surface_projection}). By adjusting its sampling time, the DLIS was able to cope with the flood of lamp light reflected off the surface. The observed intensity was more than a hundred times larger than before landing, hence the choice of background does not affect the end result. The reflectance reconstructed assuming a $1/d^2$ scaling of the SSL flux is about a third of that of the pre-landing spectra. However, when we apply the correction obtained from the DISR\#2 spare camera (see Sec.~\ref{subsec:SSL_cal}), we find the level raised to that of the others (Fig.~\ref{fig:reflectance_249}A). Close comparison of the {\bf 210} and (corrected) {\bf 249} reflectances reveals a slight mismatch between 1450 and 1500~nm (Fig.~\ref{fig:reflectance_249}B). This difference of 7\% (also present between {\bf 206} and {\bf 249}) is significant because here the responsivity is high, and the reconstruction insensitive to the choice of background (see Fig.~\ref{fig:reflectance_206_210}). We have to be careful when interpreting this deepening, though. The DISR\#3 SSL spectrum has not been measured with a target at close range, and DISR\#2 observed a drop in intensity of 5\% at 1500~nm (which is included in Fig.~\ref{fig:reflectance_249}B).

We find other absorption features in the post-landing reflectance around 1160, 1330, and 1400~nm. The 1160~nm feature is a methane band, and well modeled with a methane mixing ratio of $4.5\pm1.0$\% (Figure~\ref{fig:reflectance_249}C), assuming the DLIS is positioned 45 cm above the surface \citep{K07}. This agrees with the mixing ratio determined before landing, implying that landing did not change the methane abundance in the DLIS optical path. But where we also expect the presence of the 1400~nm methane band in the {\bf 249} reflectance, we find that it cannot be modeled satisfactory in the 1300-1450~nm range with our methane absorption model (Figure~\ref{fig:reflectance_249}C, inset). The nature of this region remains puzzling, and might even reflect imperfections in the responsivity. Assuming that the methane mixing ratio is indeed the same before and after landing leaves very little room for liquid methane. The depth of the methane absorption bands in pre-landing spectra is governed by atmospheric methane, but after landing there is so little gaseous methane present in the optical path that the spectral signature of liquid methane present on the surface would leave its mark on the absorption bands. Using the coefficients of \citet{G02} we determine an upper limit of circa 20~\textmu m for the thickness of a liquid layer on the surface, based on a fit to the 1160~nm complex.

\subsection{Full spectrum}
\label{subsec:full_spectrum}

\begin{figure}
\centering
\includegraphics[width=8cm, angle=0]{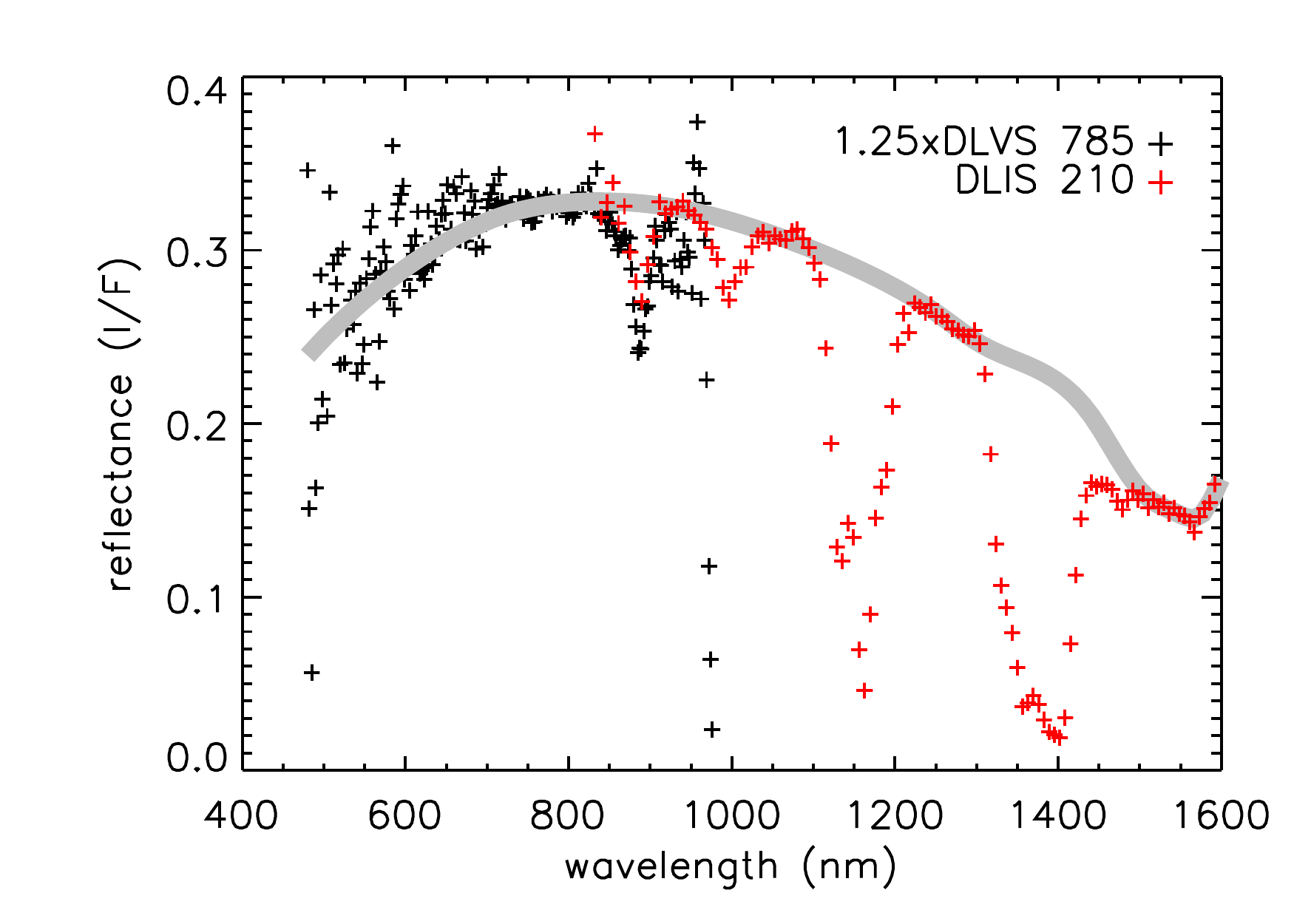}
\caption{The full reflectance spectrum of the surface around the landing site is found by scaling DLVS reflectance {\bf 785}$-1.00\times${\bf 772} to match DLIS reflectance {\bf 210}$-1.01\times${\bf 202}, and fitting a spline (thick gray line) through the methane windows. This is our best guess reflectance spectrum (for phase angle zero).}
\label{fig:albedo}
\end{figure}

In the preceding sections we have reconstructed the surface reflectance from the last DLVS and DLIS spectra before landing, making as few and simple assumptions as possible. Ideally, all pre-landing reflectances should conform, with the DLVS and DLIS agreeing at overlapping wavelengths. The two DLVS reflectances ({\bf 785} and {\bf 786}) are quite similar, as are the three DLIS reflectances ({\bf 206}, {\bf 210}, and {\bf 249}). This suggests that the method of reflectance reconstruction by scaling the SSL flux is sound, and that we successfully corrected DLIS {\bf 249} for parallax effects. But the DLVS reflectances are somewhat lower that those found by the DLIS. Figure~\ref{fig:albedo} attempts to reconcile the results from both spectrometers. When we scale the DLVS {\bf 785} reflectance up by a factor 0.25 we find reasonable agreement with the DLIS {\bf 210} reflectance. A slight discrepancy occurs in the wavelength range of overlap, where the DLVS reflectance appears to drop and the DLIS reflectance is more or less constant (curiously, it is worse for DLVS {\bf 786}). The same trend can be observed in the reconstructed SSL spectrum (see Fig.~\ref{fig:SSL_spectrum}). If, as we suspect, there it was due to a small error in the geometric correction, it would not affect the reflectance because the lamp spectrum is divided out before the correction is applied, and we do not expect steep gradients to remain after division. If it was caused by an incorrect responsivity (i.e.\ if the true lamp spectrum follows the DLIS curve in Fig.~\ref{fig:SSL_spectrum}D instead of dipping down towards 1000~nm), it would make the DLVS reflectances bend downward even more. Note that errors affecting the reconstruction of the SSL spectrum do not necessarily affect the calibration of Titan spectra, because of the much lower temperatures involved (and, consequently, different detector responsivities). In any case, the DLVS reflectance reconstruction is affected by uncertainties that are not relevant to the DLIS, so the fault likely lies with the DLVS calibration.

\subsection{Methane abundance}

The methane mixing ratios found by fitting a methane absorption model to the various reflectance spectra in this chapter agree well. The 6$\pm$2\% mixing ratio estimated from the DLVS {\bf 785} and {\bf 786} spectra is not well constrained due to the fact that the main methane absorption band is located at the red edge of the spectrum, where it is difficult to define the underlying surface reflectance because of low responsivity. The 4$\pm$1\% ratio derived from DLIS {\bf 206} is relatively uncertain because the spectrum was acquired at high altitude, and very noisy as a result. The high S/N spectrum DLIS {\bf 210} gives us the most reliable determination of the mixing ratio: 4.5$\pm$0.5\%. The post-landing DLIS {\bf 249} ratio of 4.5$\pm$1.0\% agrees with this value, but is relatively uncertain because we can fit only one out of four methane bands. All determinations agree within the range of uncertainty, and the most reliable methane mixing ratio of 4.5$\pm$0.5\% is consistent with the $4.9\pm0.3$\% measured by Huygens' GCMS instrument \citep{N05}. Interestingly, our estimate agrees with that of \citet{T05}, who derived a methane mixing ratio of 5\% from DLIS {\bf 210} at an altitude of 21~m. Adjusting the altitude to the 24.8~m we presume is correct, would decrease their mixing ratio by 0.8\%. The authors used the \citet{S93} methane absorption coefficients, which apparently leads to overestimating the degree of absorption in the strong methane bands around 1150 and 1400~nm. Using the \citet{I06} coefficients brings their mixing ratio back in the 5\% range. Compared to the \citeauthor{S93} coefficients the \citeauthor{I06} coefficients model the 1320~nm absorption shoulder well. The detailed shape of the 1000~nm band, modeled with the \citet{K98} coefficients, is not reproduced for {\bf 210}. The validity of the \citeauthor{I06} coefficients for the DLIS wavelength range is essentially confirmed by \citet{J07} who present laboratory measurements of methane absorption under Titan conditions close to the surface. These authors find a methane mixing ratio of $5.1 \pm 0.8$\% by fitting the {\bf 210} spectrum, consistent with our findings.

\subsection{Long term variability}
\label{subsec:long_term}

\begin{figure}
\centering
\includegraphics[width=8cm,angle=0]{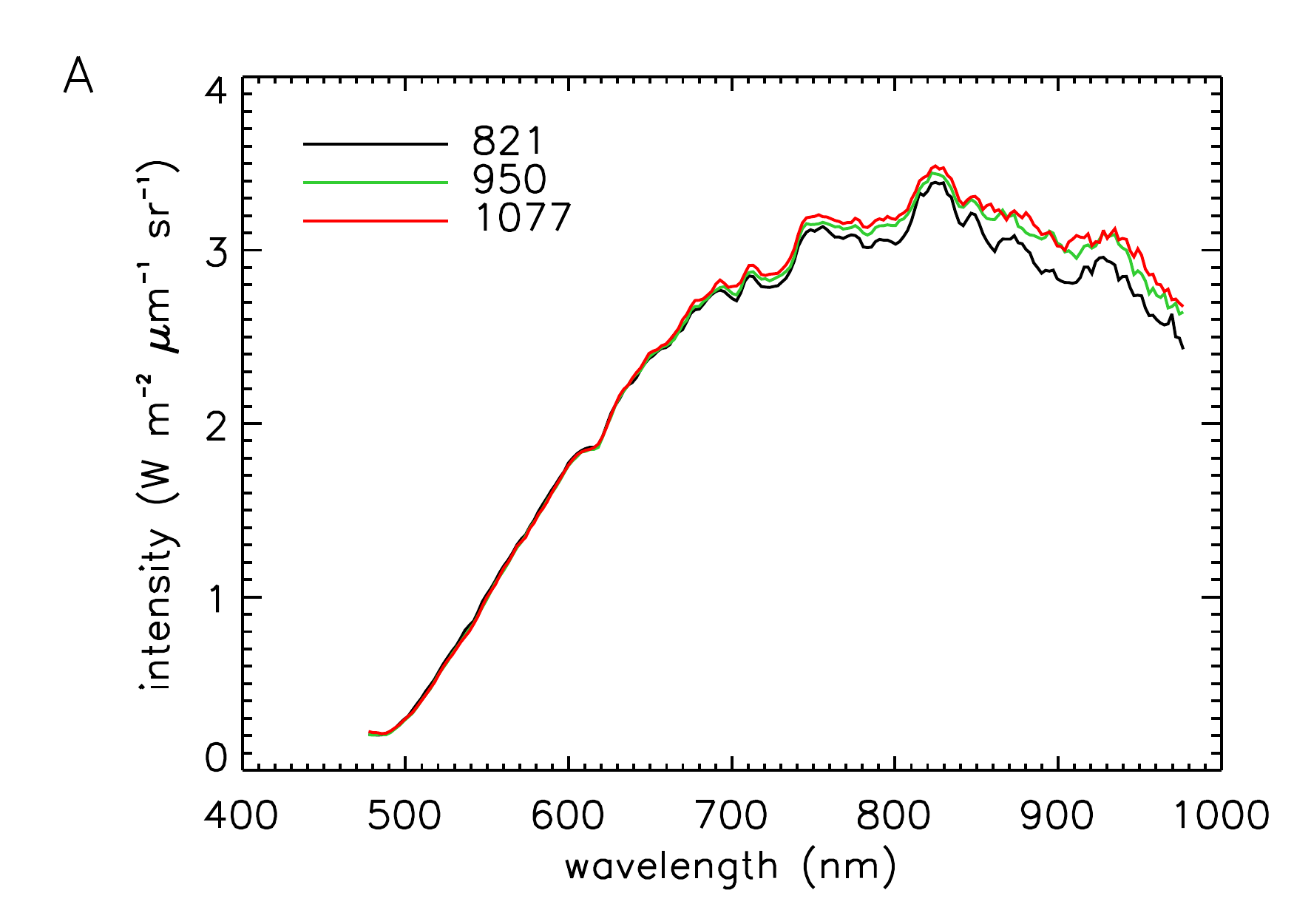}
\includegraphics[width=8cm,angle=0]{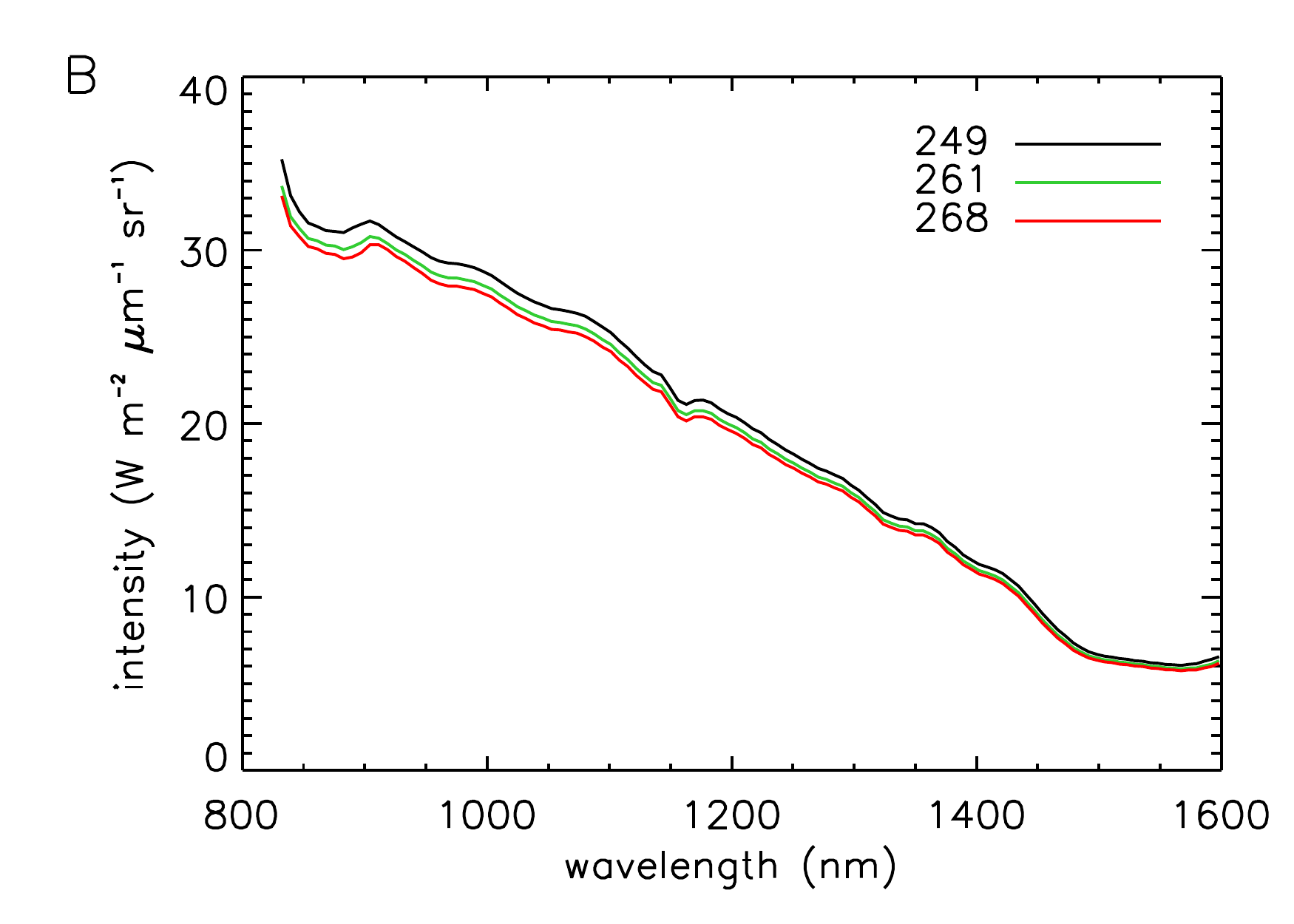}
\caption{The DISR spectrometers registered changes over the time Huygens spent on the surface. {\bf A}: DLVS {\bf 821} and {\bf 1077} are spectrum 6 (out of 8) of the first and last correctly exposed post-landing 10-column mode spectra (time elapsed 1h 8m). The increase in intensity as measured by the DLVS can be attributed to charge accumulation on the CCD. {\bf B}: DLIS {\bf 249} and {\bf 268} are the first and last correctly exposed post-landing spectra (time elapsed 1h 6m). The gradual darkening may be associated with an increase in detector temperature.} \label{fig:evolution}
\end{figure}

We investigated whether the spectra as observed by the DLVS and DLIS show any changes over the period of more than an hour in which telemetry was received from Huygens on the surface. The DLVS spectrum shows a gradual increase over time at higher wavelengths (Fig.~\ref{fig:evolution}, left). This brightness increase is not real, and can be attributed to charge accumulation on the CCD due to the strong signal. There are several lines of evidence that lead to this conclusion. First, the signal of another DISR instrument, the Downward Looking Violet photometer, which has a field of view of half a hemisphere, was constant over time. Second, the other CCD-based instruments, the imagers and the extra column which was used to gauge the amount of crosstalk, show similar behavior over time. Third, the DLVS behavior is not matched by the DLIS, which, in fact, observed a gradual darkening (Fig.~\ref{fig:evolution}, right). Whether this darkening of about 5\% is real cannot be established with certainty, but probably not either because it was not observed by any of the other downward looking instruments. It probably emerged from a switch to a different set of responsivities associated with the 14~K increase in detector temperature. However, we can still look for changes in the methane absorption bands. When we scale DLIS {\bf 261} and {\bf 268} to match {\bf 249} we find the 1160~nm methane absorption line complex to be invariant, with the implication that the methane mixing ratio was constant over time.

The presence of liquid methane on the surface at the landing site is ruled out by the post-landing DLIS spectra (see Sec.~\ref{DLIS}) and observations by other Huygens instruments \citep{F05,Z05}. However, there are hints that the subsurface is wet at a depth of several cm \citep{N05,Z05,Lo06}. Most of the energy of the 20~W SSL was deposited on an area of about 80~cm$^2$, and the heat flow downward may have heated liquid methane in the subsurface to the boiling point at 116~K (at Titan pressure). A simple simulation of the heat flow, using the parameters for a porous icy regolith from \citet{Tok05}, suggests that the top of the surface would have heated up quickly, possibly to temperatures as high as 170~K depending on the amount of free convection (this temperature assumes air flow velocities of 2~m~s$^{-1}$). Evidence for strong surface heating is found in the post-landing images in the form of changes in seeing \citep{K07}. However, due to the low thermal conductivity of this type of surface, a temperature increase of at least 20~K is reached only in the upper 2~cm of the soil. If liquid methane was present below 3~cm, evaporation would hardly have increased the mixing ratio in the air above the lamp reflection spot. While the DLIS was peering directly into the spot (see Fig.~\ref{fig:surface_projection}), the fact that it did not record any changes in the methane absorption bands does not exclude the presence of liquid methane a few centimeters below the surface.

\section{Discussion}
\label{sec:discussion}

\begin{figure}
\centering
\includegraphics[width=8cm,angle=0]{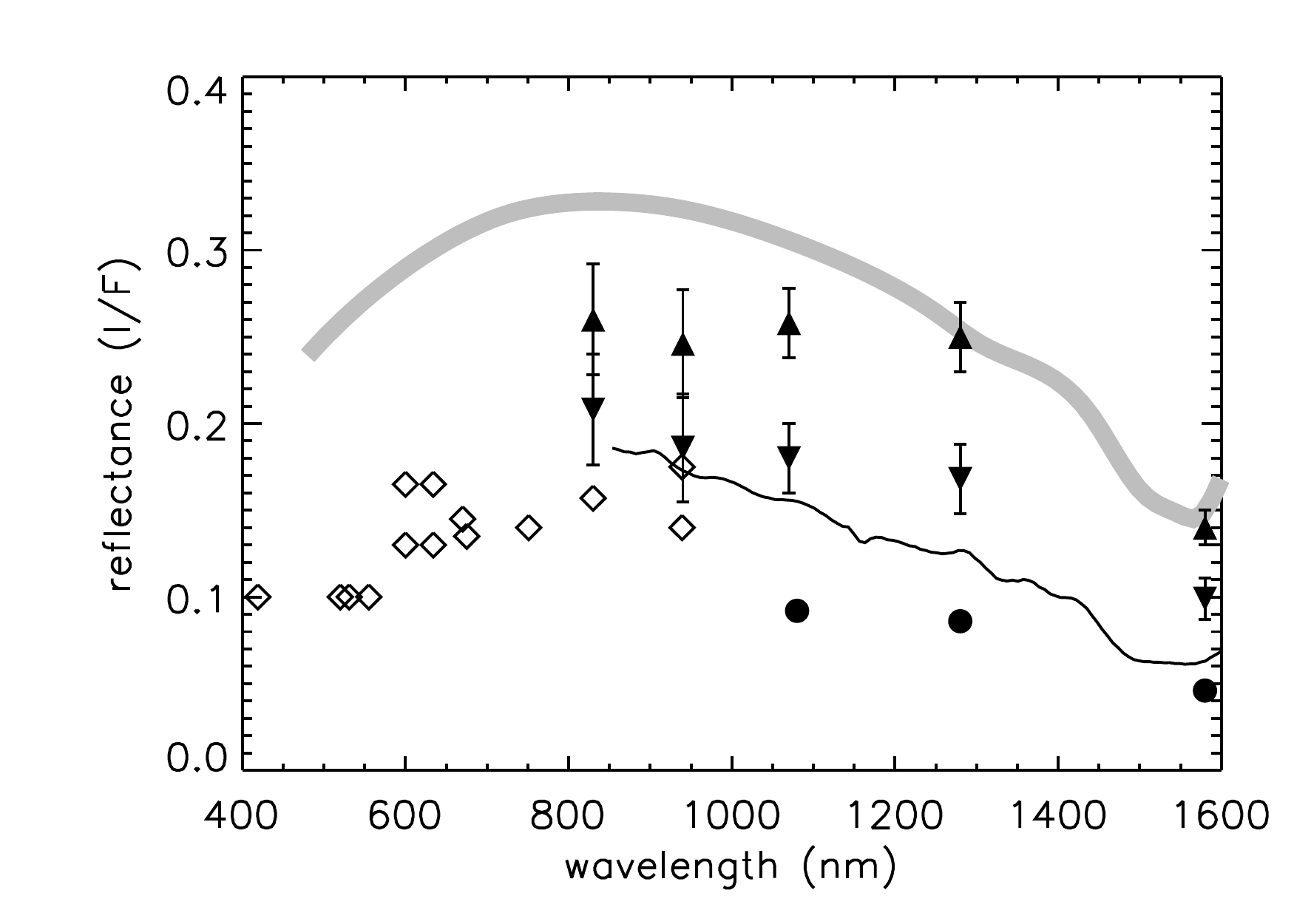}
\caption{Our best-guess reflectance spectrum from Fig.~\ref{fig:albedo} (broad gray line) compared to reflectances from the literature. Plotted are the surface albedos from the \citet{T07} atmosphere model ($\Diamond$) (L.~Doose, pers. com.), the albedos found by \citet{G03} for the leading ($\blacktriangle$) and trailing ($\blacktriangledown$) hemisphere of Titan, the albedos found by \citet{McC06} for dark equatorial terrain ($\bullet$), and the DLIS-derived reflectance from \citet{T05} (thin line).}
\label{fig:albedo_compared}
\end{figure}

Our reflectance reconstruction represents a refinement of the preliminary reconstruction by \citet{T05}. The shape of our spectrum is similar, with the DLVS part now properly calibrated and the DLIS part slightly improved. Even though the DLVS reconstruction is fraught with uncertainties we believe we find sufficient evidence for the presence of a red slope in the visible. The reflectance peaks between 800 and 900~nm, beyond which it slopes down to about half the peak value at 1500~nm. This blue slope is virtually featureless with the exception of an absorption feature at 1500~nm. Whereas we find the reflectance by scaling our result to the lamp flux measured before launch, \citeauthor{T05} scale their reconstruction to the average of the ratio of the up- and downward flux, derived from seven low-altitude DLIS and ULIS spectra (recently confirmed by \citealt{J07}). Figure~\ref{fig:albedo_compared} shows that our approach leads to dramatically higher values of the overall reflectance.

How does our best-guess reflectance spectrum compare to those found by other teams? \citet{G03} determined the surface albedo in near-IR methane windows of the leading and trailing side of Titan by means of a radiative transfer model. The leading hemisphere features the bright `continent' Xanadu, and therefore has a higher average albedo than the trailing hemisphere, which has dark terrain distributed around the equator \citep{P05}. Huygens' landing site is close to the equator on the trailing hemisphere, but the lake terrain observed by DISR is not covered by the very dark dunes which are ubiquitous in the equatorial dark terrain \citep{L06}. Hence, the albedo of the landing site must be lower than that of the trailing hemisphere, but higher than that of the dune-covered terrain \citep[reconstructed by][]{McC06}. Figure~\ref{fig:albedo_compared} shows that our reflectance spectrum does not meet this requirement; it is higher than the others, comparing only to that of Titan's leading hemisphere. The \citet{T05} reflectance better meets the expectations. Recently, \citet{T07} determined the visual part of the reflectance of the Huygens landing site by means of a comprehensive atmosphere model. Their results (Fig.~\ref{fig:albedo_compared}, diamonds) confirm the low reflectance of the landing site, and we must accept that the discrepancy with ours is real.

The overall reflectance of our reconstruction is higher over the full wavelength range than that determined by all other teams. Does our reconstruction fail? If there are problems associated with scaling the lamp flux and/or parallax effects, the reflectances derived from the two DLVS and three DLIS spectra would disagree more. Perhaps our altitude scale is incorrect. We use the SSP landing velocity of 4.60~m~s$^{-1}$ (\citealt{Z05}, confirmed by \citealt{TG06}) to compute the altitude of the last observations before landing. Using the HASI velocity of 4.33~m~s$^{-1}$ \citep{F05} would decrease our reflectance only by 12\%. Other unlikely explanations range from an unidentified problem with the SSL calibration to the last three footprints covering unusually bright terrain. There is no evidence for either. The most likely explanation has to do with the phase angle at which the observations were acquired. If the surface of Titan at the landing site exhibits an opposition effect (a strong increase in brightness towards zero phase angle), e.g.\ through a combination of shadow hiding and coherent backscatter \citep{H81,H02}, our reflectance can be much higher than those of other groups because the phase angle of the last pre-landing spectra were less than a degree. We can calculate the phase angles from the size of the SSL window (5.2~cm diameter) and the distance of its center to the spectrometer windows on either side (DLIS: 3.2~cm, DLVS: 6.8~cm). DLIS spectra {\bf 206} and {\bf 210}, associated with the highest overall reflectance, were acquired at the smallest phase angles: $0.031^\circ \pm 0.025^\circ$ and $0.069^\circ \pm 0.056^\circ$, respectively. The phase angle of DLVS {\bf 786} is higher ($0.45^\circ \pm 0.17^\circ$) and its reflectance lower ($\sim$20\%) than that of DLIS {\bf 210}, as would be expected in case of an opposition effect. The phase angle of DLVS {\bf 785} is intermediate to the previous two at $0.23^\circ \pm 0.09^\circ$, and we would expect its reflectance to be intermediate as well, contrary to what we observe. However, parallax would have pushed the lamp beam and the DLVS footprint away from each other (see Fig.~\ref{fig:cal_exp_slits}), and the true SSL flux must have been lower in reality than that used, which makes the reconstructed {\bf 785} reflectance an underestimate. Post-landing DLIS spectrum {\bf 249} observed the surface over a wide range of phase angles (approximately $0.70^\circ$-$6.7^\circ$), so the associated reflectance is expected to be lower than that of {\bf 210}. The fact that it is not could imply either surface heterogeneity (as seen in the post-landing images), or that DISR\#2 is not sufficiently similar to DISR\#3.

Albedos from other workers have all been derived from observations that were significantly affected by the atmosphere. For example, the ratio method employed by \citet{T05} averaged eight DLIS observations of reflected sunlight at phase angles ranging from $13^\circ$ to $57^\circ$. But the diffuse nature of Titan surface illumination would subdue shadow hiding, the dominant mechanism at these phase angles. Zero phase angle observations from outside Titan's atmosphere would not register an increased reflectance, because even though the surface is visible through the near-IR methane windows, hardly any direct sunlight reaches it, which is an essential prerequisite for coherent backscatter. Thus, we suggest that our high reflectance represents the opposition brightness surge, the apparent strength of which is within reasonable bounds. Opposition surges of similar magnitude have been observed for the surfaces of other low-albedo solar system bodies, like the Moon \citep{P69}, and Callisto \citep{DV97}. We explore this topic further in an upcoming paper. Note that while we consider the opposition effect to be a natural explanation for our high reflectance, we cannot exclude the possibility that Huygens landed on a bright ridge, since the landing site was not imaged in detail (see Fig.~\ref{fig:MNS&VLNS_projection}).

The reflectance spectrum offers clues about the surface composition at Huygens' landing site. What spectral signatures do we expect? Titan is large enough to be differentiated, so presumably the rocks we see in the surface images are ice cobbles. Organic ices, carbon dioxide, and ammonia may be mixed in with the water ice. Aerosols that continuously rain out of Titan's haze likely cover all of this, but the presence of cobbles indicates that it does not form a thick uniform blanket. Whether the aerosols adhere to the rocks is not clear; they are thought to `harden' on their way to the surface \citep{D02}. If so, they may behave like dust, and wind may blow them off the cobbles to expose clean water ice. As to the interpretation of the reflectance spectrum, we agree with \citet{T05}. We briefly summarize the main issues. The red slope in the visible is characteristic for complex organic material. The blue slope in the near-IR on the other hand, has not yet been associated with any know combination of ices and organic material. It is featureless except for a single absorption feature at 1500~nm, and does not show lines of organic ices, silicates, or ammonia. The nature of the 1500~nm feature remains uncertain. An absorption band at this wavelength is present in both water ice \citep{C81b} and tholin spectra \citep{C91,B06}, but an interpretation in terms of either material is problematic. Tholins are red over the whole DISR wavelength range, which is inconsistent with the observed blue slope. In water ice spectra that feature a blue slope the 1500~nm absorption band is deep and the accompanying 1040 and 1250~nm overtone bands are clearly visible, whereas spectra with a 1500~nm band as shallow as that observed are flat. Noteworthy is the observation that this feature deepens after landing. The implications of this are not clear. Either the act of landing made this feature deeper (dust thrown up revealing fresh water ice?), or the surface is heterogeneous at small scales, a notion supported by the post-landing images.

\section{Conclusions}
\label{sec:conclusions}

We have reconstructed the reflectance spectrum of Titan's surface at the Huygens landing site from pre-landing DLVS and DLIS spectra that show the presence of lamp light. Our reconstruction improves upon the preliminary reconstruction by \citet{T05}, and may serve as a reference for those who wish to constrain the surface composition by modeling.

The reflectance spectrum features a red slope in the visible, confirming the presence of complex organics, and an almost featureless blue slope in the near-IR, which, so far, has defied interpretation. The only unambiguous absorption feature is at 1500~nm, which may be associated with either tholins or water ice. We agree with \citet{T05} that the evidence for water ice in DISR spectra is inconclusive. We find this line to be significantly deeper after landing, possibly indicating surface heterogeneity.

Our reconstruction method is a direct one, depending on few assumptions. More specifically, it does not depend on radiative transfer model calculations. We find a almost twofold brightening of the surface reflectance at phase angle zero compared to that found by other methods. We present this as evidence for an opposition surge.

We find that the \citet{I06} coefficients model the near-IR methane absorption bands slightly better than those of \citet{S93}. The \citet{K98} coefficients model the spectrum below 1050~nm well, albeit at a methane abundance a third higher than what follows from the \citeauthor{I06} coefficients. We conclude that the atmospheric methane mixing ratio as determined from pre- and post-landing spectra is $4.5 \pm 0.5$\%, consistent with the Huygens' GCMS value. The post-landing DLIS spectra do not support the presence of a layer of liquid methane on the surface of the landing site, but do not rule out a wet soil at a depth of a few centimeters.

\section*{Acknowledgements}

The authors are grateful for the support provided by M.~Tomasko and his team at LPL, especially C.~See for assisting in the camera spare experiments. We also thank B.~Schmitt for measuring the calibration target reflectances, E.~Lellouch for providing the water absorption coefficients, and E.~Karkoschka for reviewing the manuscript. In addition, we thank B.~B\'ezard and an anonymous reviewer for excellent comments that led to a significantly improved paper. We acknowledge funding by the {\em Deutsches Zentrum f\"ur Luft und Raumfahrt} (DLR) through grant 50 OH 98044.

\bibliography{reflectance}

\begin{thebibliography}{40}
\expandafter\ifx\csname natexlab\endcsname\relax\def\natexlab#1{#1}\fi
\expandafter\ifx\csname url\endcsname\relax
  \def\url#1{\texttt{#1}}\fi
\expandafter\ifx\csname urlprefix\endcsname\relax\def\urlprefix{URL }\fi

\bibitem[{{Atreya} et~al.(2006){Atreya}, {Adams}, {Niemann},
  {Demick-Montelara}, {Owen}, {Fulchignoni}, {Ferri}, and {Wilson}}]{A06}
{Atreya}, S.~K., {Adams}, E.~Y., {Niemann}, H.~B., {Demick-Montelara}, J.~E.,
  {Owen}, T.~C., {Fulchignoni}, M., {Ferri}, F., {Wilson}, E.~H., Oct. 2006.
  {Titan's methane cycle}. \planss 54, 1177--1187.

\bibitem[{{Bernard} et~al.(2006){Bernard}, {Quirico}, {Brissaud}, {Montagnac},
  {Reynard}, {McMillan}, {Coll}, {Nguyen}, {Raulin}, and {Schmitt}}]{B06}
{Bernard}, J.-M., {Quirico}, E., {Brissaud}, O., {Montagnac}, G., {Reynard},
  B., {McMillan}, P., {Coll}, P., {Nguyen}, M.-J., {Raulin}, F., {Schmitt}, B.,
  Nov. 2006. {Reflectance spectra and chemical structure of Titan's tholins:
  Application to the analysis of Cassini-Huygens observations}. Icarus 185,
  301--307.

\bibitem[{{Clark}(1981)}]{C81b}
{Clark}, R.~N., Apr. 1981. {Water frost and ice: The near-infrared spectral
  reflectance 0.65-2.5 $\mu$m}. \jgr 86, 3087--3096.

\bibitem[{{Coustenis} et~al.(1995){Coustenis}, {Lellouch}, {Maillard}, and
  {McKay}}]{C95}
{Coustenis}, A., {Lellouch}, E., {Maillard}, J.~P., {McKay}, C.~P., Nov. 1995.
  {Titan's Surface: Composition and Variability from the Near-Infrared Albedo.}
  Icarus 118, 87--104.

\bibitem[{{Cruikshank} et~al.(1991){Cruikshank}, {Allamandola}, {Hartmann},
  {Tholen}, {Brown}, {Matthews}, and {Bell}}]{C91}
{Cruikshank}, D.~P., {Allamandola}, L.~J., {Hartmann}, W.~K., {Tholen}, D.~J.,
  {Brown}, R.~H., {Matthews}, C.~N., {Bell}, J.~F., Dec. 1991. {Solid
  C$\equiv$N Bearing Material on Outer Solar System Bodies}. Icarus 94,
  345--353.

\bibitem[{{Dimitrov} and {Bar-Nun}(2002)}]{D02}
{Dimitrov}, V., {Bar-Nun}, A., Apr. 2002. {Aging of Titan's Aerosols}. Icarus
  156, 530--538.

\bibitem[{{Domingue} and {Verbiscer}(1997)}]{DV97}
{Domingue}, D., {Verbiscer}, A., Jul. 1997. {Re-Analysis of the Solar Phase
  Curves of the Icy Galilean Satellites}. Icarus 128, 49--74.

\bibitem[{{Elachi} et~al.(2005){Elachi}, {Wall}, {Allison}, {Anderson},
  {Boehmer}, {Callahan}, {Encrenaz}, {Flamini}, {Franceschetti}, {Gim},
  {Hamilton}, {Hensley}, {Janssen}, {Johnson}, {Kelleher}, {Kirk}, {Lopes},
  {Lorenz}, {Lunine}, {Muhleman}, {Ostro}, {Paganelli}, {Picardi}, {Posa},
  {Roth}, {Seu}, {Shaffer}, {Soderblom}, {Stiles}, {Stofan}, {Vetrella},
  {West}, {Wood}, {Wye}, and {Zebker}}]{E05}
{Elachi}, C., {Wall}, S., {Allison}, M., {Anderson}, Y., {Boehmer}, R.,
  {Callahan}, P., {Encrenaz}, P., {Flamini}, E., {Franceschetti}, G., {Gim},
  Y., {Hamilton}, G., {Hensley}, S., {Janssen}, M., {Johnson}, W., {Kelleher},
  K., {Kirk}, R., {Lopes}, R., {Lorenz}, R., {Lunine}, J., {Muhleman}, D.,
  {Ostro}, S., {Paganelli}, F., {Picardi}, G., {Posa}, F., {Roth}, L., {Seu},
  R., {Shaffer}, S., {Soderblom}, L., {Stiles}, B., {Stofan}, E., {Vetrella},
  S., {West}, R., {Wood}, C., {Wye}, L., {Zebker}, H., May 2005. {Cassini Radar
  Views the Surface of Titan}. Science 308, 970--974.

\bibitem[{{Fulchignoni} et~al.(2005){Fulchignoni}, {Ferri}, {Angrilli}, {Ball},
  {Bar-Nun}, {Barucci}, {Bettanini}, {Bianchini}, {Borucki}, {Colombatti},
  {Coradini}, {Coustenis}, {Debei}, {Falkner}, {Fanti}, {Flamini}, {Gaborit},
  {Grard}, {Hamelin}, {Harri}, {Hathi}, {Jernej}, {Leese}, {Lehto}, {Lion
  Stoppato}, {L{\'o}pez-Moreno}, {M{\"a}kinen}, {McDonnell}, {McKay},
  {Molina-Cuberos}, {Neubauer}, {Pirronello}, {Rodrigo}, {Saggin},
  {Schwingenschuh}, {Seiff}, {Sim{\~o}es}, {Svedhem}, {Tokano}, {Towner},
  {Trautner}, {Withers}, and {Zarnecki}}]{F05}
{Fulchignoni}, M., {Ferri}, F., {Angrilli}, F., {Ball}, A.~J., {Bar-Nun}, A.,
  {Barucci}, M.~A., {Bettanini}, C., {Bianchini}, G., {Borucki}, W.,
  {Colombatti}, G., {Coradini}, M., {Coustenis}, A., {Debei}, S., {Falkner},
  P., {Fanti}, G., {Flamini}, E., {Gaborit}, V., {Grard}, R., {Hamelin}, M.,
  {Harri}, A.~M., {Hathi}, B., {Jernej}, I., {Leese}, M.~R., {Lehto}, A., {Lion
  Stoppato}, P.~F., {L{\'o}pez-Moreno}, J.~J., {M{\"a}kinen}, T., {McDonnell},
  J.~A.~M., {McKay}, C.~P., {Molina-Cuberos}, G., {Neubauer}, F.~M.,
  {Pirronello}, V., {Rodrigo}, R., {Saggin}, B., {Schwingenschuh}, K., {Seiff},
  A., {Sim{\~o}es}, F., {Svedhem}, H., {Tokano}, T., {Towner}, M.~C.,
  {Trautner}, R., {Withers}, P., {Zarnecki}, J.~C., Dec. 2005. {In situ
  measurements of the physical characteristics of Titan's environment}. \nat
  438, 785--791.

\bibitem[{{Griffith} et~al.(2003){Griffith}, {Owen}, {Geballe}, {Rayner}, and
  {Rannou}}]{G03}
{Griffith}, C.~A., {Owen}, T., {Geballe}, T.~R., {Rayner}, J., {Rannou}, P.,
  Apr. 2003. {Evidence for the Exposure of Water Ice on Titan's Surface}.
  Science 300, 628--630.

\bibitem[{{Grundy} et~al.(2002){Grundy}, {Schmitt}, and {Quirico}}]{G02}
{Grundy}, W.~M., {Schmitt}, B., {Quirico}, E., Feb. 2002. {The
  Temperature-Dependent Spectrum of Methane Ice I between 0.7 and 5 {$\mu$}m
  and Opportunities for Near-Infrared Remote Thermometry}. Icarus 155,
  486--496.

\bibitem[{{Hapke}(1981)}]{H81}
{Hapke}, B., Apr. 1981. {Bidirectional reflectance spectroscopy. I - Theory}.
  \jgr 86, 3039--3054.

\bibitem[{{Hapke}(2002)}]{H02}
{Hapke}, B., Jun. 2002. {Bidirectional Reflectance Spectroscopy 5. The Coherent
  Backscatter Opposition Effect and Anisotropic Scattering}. Icarus 157,
  523--534.

\bibitem[{{Irwin} et~al.(2006){Irwin}, {Sromovsky}, {Strong}, {Sihra},
  {Teanby}, {Bowles}, {Calcutt}, and {Remedios}}]{I06}
{Irwin}, P.~G.~J., {Sromovsky}, L.~A., {Strong}, E.~K., {Sihra}, K., {Teanby},
  N.~A., {Bowles}, N., {Calcutt}, S.~B., {Remedios}, J.~J., Mar. 2006.
  {Improved near-infrared methane band models and k-distribution parameters
  from 2000 to 9500 cm$^{-1}$ and implications for interpretation of outer
  planet spectra}. Icarus 181, 309--319.

\bibitem[{{Jacquemart} et~al.(2008){Jacquemart}, {Lellouch}, {B{\'e}zard}, {de
  Bergh}, {Coustenis}, {Lacome}, {Schmitt}, and {Tomasko}}]{J07}
{Jacquemart}, D., {Lellouch}, E., {B{\'e}zard}, B., {de Bergh}, C.,
  {Coustenis}, A., {Lacome}, N., {Schmitt}, B., {Tomasko}, M., Apr. 2008. {New
  laboratory measurements of CH $_{4}$ in Titan's conditions and a reanalysis
  of the DISR near-surface spectra at the Huygens landing site}. \planss 56,
  613--623.

\bibitem[{{Karkoschka}(1998)}]{K98}
{Karkoschka}, E., May 1998. {Methane, Ammonia, and Temperature Measurements of
  the Jovian Planets and Titan from CCD-Spectrophotometry}. Icarus 133,
  134--146.

\bibitem[{{Karkoschka} et~al.(2007){Karkoschka}, {Tomasko}, {Doose}, {See},
  {McFarlane}, {Schr{\"o}der}, and {Rizk}}]{K07}
{Karkoschka}, E., {Tomasko}, M.~G., {Doose}, L.~R., {See}, C., {McFarlane},
  E.~A., {Schr{\"o}der}, S.~E., {Rizk}, B., Nov. 2007. {DISR imaging and the
  geometry of the descent of the Huygens probe within Titan's atmosphere}.
  \planss 55, 1896--1935.

\bibitem[{{Keller} et~al.(2008){Keller}, {Grieger}, {K{\"u}ppers},
  {Schr{\"o}der}, {Skorov}, and {Tomasko}}]{MPS07}
{Keller}, H.~U., {Grieger}, B., {K{\"u}ppers}, M., {Schr{\"o}der}, S.~E.,
  {Skorov}, Y.~V., {Tomasko}, M.~G., Apr. 2008. {The properties of Titan's
  surface at the Huygens landing site from DISR observations}. \planss 56,
  728--752.

\bibitem[{{Lellouch} et~al.(2004){Lellouch}, {Schmitt}, {Coustenis}, and
  {Cuby}}]{L04}
{Lellouch}, E., {Schmitt}, B., {Coustenis}, A., {Cuby}, J.-G., Mar. 2004.
  {Titan's 5-micron lightcurve}. Icarus 168, 209--214.

\bibitem[{{Lorenz} et~al.(2006{\natexlab{a}}){Lorenz}, {Niemann}, {Harpold},
  {Way}, and {Zarnecki}}]{Lo06}
{Lorenz}, R.~D., {Niemann}, H.~B., {Harpold}, D., {Way}, S., {Zarnecki}, J.~C.,
  Nov. 2006{\natexlab{a}}. {Titan's damp ground: Constraints on Titan surface
  thermal properties from the temperature evolution of the Huygens GCMS inlet}.
  Meteoritics \& Planetary Science 41, 1705--1714.

\bibitem[{{Lorenz} et~al.(2006{\natexlab{b}}){Lorenz}, {Wall}, {Radebaugh},
  {Boubin}, {Reffet}, {Janssen}, {Stofan}, {Lopes}, {Kirk}, {Elachi}, {Lunine},
  {Mitchell}, {Paganelli}, {Soderblom}, {Wood}, {Wye}, {Zebker}, {Anderson},
  {Ostro}, {Allison}, {Boehmer}, {Callahan}, {Encrenaz}, {Ori}, {Francescetti},
  {Gim}, {Hamilton}, {Hensley}, {Johnson}, {Kelleher}, {Muhleman}, {Picardi},
  {Posa}, {Roth}, {Seu}, {Shaffer}, {Stiles}, {Vetrella}, {Flamini}, and
  {West}}]{L06}
{Lorenz}, R.~D., {Wall}, S., {Radebaugh}, J., {Boubin}, G., {Reffet}, E.,
  {Janssen}, M., {Stofan}, E., {Lopes}, R., {Kirk}, R., {Elachi}, C., {Lunine},
  J., {Mitchell}, K., {Paganelli}, F., {Soderblom}, L., {Wood}, C., {Wye}, L.,
  {Zebker}, H., {Anderson}, Y., {Ostro}, S., {Allison}, M., {Boehmer}, R.,
  {Callahan}, P., {Encrenaz}, P., {Ori}, G.~G., {Francescetti}, G., {Gim}, Y.,
  {Hamilton}, G., {Hensley}, S., {Johnson}, W., {Kelleher}, K., {Muhleman}, D.,
  {Picardi}, G., {Posa}, F., {Roth}, L., {Seu}, R., {Shaffer}, S., {Stiles},
  B., {Vetrella}, S., {Flamini}, E., {West}, R., May 2006{\natexlab{b}}. {The
  Sand Seas of Titan: Cassini RADAR Observations of Longitudinal Dunes}.
  Science 312, 724--727.

\bibitem[{{Lunine} et~al.(2008){Lunine}, {Elachi}, {Wall}, {Janssen},
  {Allison}, {Anderson}, {Boehmer}, {Callahan}, {Encrenaz}, {Flamini},
  {Franceschetti}, {Gim}, {Hamilton}, {Hensley}, {Johnson}, {Kelleher}, {Kirk},
  {Lopes}, {Lorenz}, {Muhleman}, {Orosei}, {Ostro}, {Paganelli}, {Paillou},
  {Picardi}, {Posa}, {Radebaugh}, {Roth}, {Seu}, {Shaffer}, {Soderblom},
  {Stiles}, {Stofan}, {Vetrella}, {West}, {Wood}, {Wye}, {Zebker}, {Alberti},
  {Karkoschka}, {Rizk}, {McFarlane}, {See}, and {Kazeminejad}}]{LE07}
{Lunine}, J.~I., {Elachi}, C., {Wall}, S.~D., {Janssen}, M.~A., {Allison},
  M.~D., {Anderson}, Y., {Boehmer}, R., {Callahan}, P., {Encrenaz}, P.,
  {Flamini}, E., {Franceschetti}, G., {Gim}, Y., {Hamilton}, G., {Hensley}, S.,
  {Johnson}, W.~T.~K., {Kelleher}, K., {Kirk}, R.~L., {Lopes}, R.~M., {Lorenz},
  R., {Muhleman}, D.~O., {Orosei}, R., {Ostro}, S.~J., {Paganelli}, F.,
  {Paillou}, P., {Picardi}, G., {Posa}, F., {Radebaugh}, J., {Roth}, L.~E.,
  {Seu}, R., {Shaffer}, S., {Soderblom}, L.~A., {Stiles}, B., {Stofan}, E.~R.,
  {Vetrella}, S., {West}, R., {Wood}, C.~A., {Wye}, L., {Zebker}, H.,
  {Alberti}, G., {Karkoschka}, E., {Rizk}, B., {McFarlane}, E., {See}, C.,
  {Kazeminejad}, B., May 2008. {Titan's diverse landscapes as evidenced by
  Cassini RADAR's third and fourth looks at Titan}. \icarus 195, 415--433.

\bibitem[{{Lunine} et~al.(1983){Lunine}, {Stevenson}, and {Yung}}]{L83}
{Lunine}, J.~I., {Stevenson}, D.~J., {Yung}, Y.~L., Dec. 1983. {Ethane Ocean on
  Titan}. Science 222, 1229--1230.

\bibitem[{{McCord} et~al.(2006){McCord}, {Hansen}, {Buratti}, {Clark},
  {Cruikshank}, {D'Aversa}, {Griffith}, {Baines}, {Brown}, {Dalle Ore},
  {Filacchione}, {Formisano}, {Hibbitts}, {Jaumann}, {Lunine}, {Nelson},
  {Sotin}, and {the Cassini VIMS Team}}]{McC06}
{McCord}, T.~B., {Hansen}, G.~B., {Buratti}, B.~J., {Clark}, R.~N.,
  {Cruikshank}, D.~P., {D'Aversa}, E., {Griffith}, C.~A., {Baines}, E.~K.~H.,
  {Brown}, R.~H., {Dalle Ore}, C.~M., {Filacchione}, G., {Formisano}, V.,
  {Hibbitts}, C.~A., {Jaumann}, R., {Lunine}, J.~I., {Nelson}, R.~M., {Sotin},
  C., {the Cassini VIMS Team}, Dec. 2006. {Composition of Titan's surface from
  Cassini VIMS}. \planss 54, 1524--1539.

\bibitem[{{Niemann} et~al.(2005){Niemann}, {Atreya}, {Bauer}, {Carignan},
  {Demick}, {Frost}, {Gautier}, {Haberman}, {Harpold}, {Hunten}, {Israel},
  {Lunine}, {Kasprzak}, {Owen}, {Paulkovich}, {Raulin}, {Raaen}, and
  {Way}}]{N05}
{Niemann}, H.~B., {Atreya}, S.~K., {Bauer}, S.~J., {Carignan}, G.~R., {Demick},
  J.~E., {Frost}, R.~L., {Gautier}, D., {Haberman}, J.~A., {Harpold}, D.~N.,
  {Hunten}, D.~M., {Israel}, G., {Lunine}, J.~I., {Kasprzak}, W.~T., {Owen},
  T.~C., {Paulkovich}, M., {Raulin}, F., {Raaen}, E., {Way}, S.~H., Dec. 2005.
  {The abundances of constituents of Titan's atmosphere from the GCMS
  instrument on the Huygens probe}. \nat 438, 779--784.

\bibitem[{{Pohn} et~al.(1969){Pohn}, {Radin}, and {Wildey}}]{P69}
{Pohn}, H.~A., {Radin}, H.~W., {Wildey}, R.~L., Sep. 1969. {The Moon's
  photometric function near zero phase angle from Apollo 8 photography}. \apjl
  157, L193--L195.

\bibitem[{{Porco} et~al.(2005){Porco}, {Baker}, {Barbara}, {Beurle}, {Brahic},
  {Burns}, {Charnoz}, {Cooper}, {Dawson}, {Del Genio}, {Denk}, {Dones},
  {Dyudina}, {Evans}, {Fussner}, {Giese}, {Grazier}, {Helfenstein},
  {Ingersoll}, {Jacobson}, {Johnson}, {McEwen}, {Murray}, {Neukum}, {Owen},
  {Perry}, {Roatsch}, {Spitale}, {Squyres}, {Thomas}, {Tiscareno}, {Turtle},
  {Vasavada}, {Veverka}, {Wagner}, and {West}}]{P05}
{Porco}, C.~C., {Baker}, E., {Barbara}, J., {Beurle}, K., {Brahic}, A.,
  {Burns}, J.~A., {Charnoz}, S., {Cooper}, N., {Dawson}, D.~D., {Del Genio},
  A.~D., {Denk}, T., {Dones}, L., {Dyudina}, U., {Evans}, M.~W., {Fussner}, S.,
  {Giese}, B., {Grazier}, K., {Helfenstein}, P., {Ingersoll}, A.~P.,
  {Jacobson}, R.~A., {Johnson}, T.~V., {McEwen}, A., {Murray}, C.~D., {Neukum},
  G., {Owen}, W.~M., {Perry}, J., {Roatsch}, T., {Spitale}, J., {Squyres}, S.,
  {Thomas}, P., {Tiscareno}, M., {Turtle}, E.~P., {Vasavada}, A.~R., {Veverka},
  J., {Wagner}, R., {West}, R., Mar. 2005. {Imaging of Titan from the Cassini
  spacecraft}. \nat 434, 159--168.

\bibitem[{{Sagan} and {Dermott}(1982)}]{S82}
{Sagan}, C., {Dermott}, S.~F., Dec. 1982. {The tide in the seas of Titan}. \nat
  300, 731--733.

\bibitem[{{Sagan} and {Khare}(1979)}]{C79}
{Sagan}, C., {Khare}, B.~N., Jan. 1979. {Tholins - Organic chemistry of
  interstellar grains and gas}. \nat 277, 102--107.

\bibitem[{{Schr\"oder}(2007)}]{SES07}
{Schr\"oder}, S.~E., 2007. {Investigating the Surface of Titan with the Descent
  Imager/Spectral Radiometer onboard Huygens}. Ph.D. thesis, {Universit\"at
  G\"ottingen, Germany}.

\bibitem[{{Stofan} et~al.(2007){Stofan}, {Elachi}, {Lunine}, {Lorenz},
  {Stiles}, {Mitchell}, {Ostro}, {Soderblom}, {Wood}, {Zebker}, {Wall},
  {Janssen}, {Kirk}, {Lopes}, {Paganelli}, {Radebaugh}, {Wye}, {Anderson},
  {Allison}, {Boehmer}, {Callahan}, {Encrenaz}, {Flamini}, {Francescetti},
  {Gim}, {Hamilton}, {Hensley}, {Johnson}, {Kelleher}, {Muhleman}, {Paillou},
  {Picardi}, {Posa}, {Roth}, {Seu}, {Shaffer}, {Vetrella}, and {West}}]{S07}
{Stofan}, E.~R., {Elachi}, C., {Lunine}, J.~I., {Lorenz}, R.~D., {Stiles}, B.,
  {Mitchell}, K.~L., {Ostro}, S., {Soderblom}, L., {Wood}, C., {Zebker}, H.,
  {Wall}, S., {Janssen}, M., {Kirk}, R., {Lopes}, R., {Paganelli}, F.,
  {Radebaugh}, J., {Wye}, L., {Anderson}, Y., {Allison}, M., {Boehmer}, R.,
  {Callahan}, P., {Encrenaz}, P., {Flamini}, E., {Francescetti}, G., {Gim}, Y.,
  {Hamilton}, G., {Hensley}, S., {Johnson}, W.~T.~K., {Kelleher}, K.,
  {Muhleman}, D., {Paillou}, P., {Picardi}, G., {Posa}, F., {Roth}, L., {Seu},
  R., {Shaffer}, S., {Vetrella}, S., {West}, R., Jan. 2007. {The lakes of
  Titan}. \nat 445, 61--64.

\bibitem[{{Strong} et~al.(1993){Strong}, {Taylor}, {Calcutt}, {Remedios}, and
  {Ballard}}]{S93}
{Strong}, K., {Taylor}, F.~W., {Calcutt}, S.~B., {Remedios}, J.~J., {Ballard},
  J., 1993. {Spectral parameters of self- and hydrogen-broadened methane from
  2000 to 9500 cm$^{-1}$ for remote sounding of the atmosphere of Jupiter}.
  Journal of Quantitative Spectroscopy and Radiative Transfer 50, 363--429.

\bibitem[{{Tobie} et~al.(2006){Tobie}, {Lunine}, and {Sotin}}]{T06}
{Tobie}, G., {Lunine}, J.~I., {Sotin}, C., Mar. 2006. {Episodic outgassing as
  the origin of atmospheric methane on Titan}. \nat 440, 61--64.

\bibitem[{{Tokano}(2005)}]{Tok05}
{Tokano}, T., Jan. 2005. {Meteorological assessment of the surface temperatures
  on Titan: constraints on the surface type}. Icarus 173, 222--242.

\bibitem[{{Tomasko} et~al.(2005){Tomasko}, {Archinal}, {Becker}, {B{\'e}zard},
  {Bushroe}, {Combes}, {Cook}, {Coustenis}, {de Bergh}, {Dafoe}, {Doose},
  {Dout{\'e}}, {Eibl}, {Engel}, {Gliem}, {Grieger}, {Holso}, {Howington-Kraus},
  {Karkoschka}, {Keller}, {Kirk}, {Kramm}, {K{\"u}ppers}, {Lanagan},
  {Lellouch}, {Lemmon}, {Lunine}, {McFarlane}, {Moores}, {Prout}, {Rizk},
  {Rosiek}, {Rueffer}, {Schr{\"o}der}, {Schmitt}, {See}, {Smith}, {Soderblom},
  {Thomas}, and {West}}]{T05}
{Tomasko}, M.~G., {Archinal}, B., {Becker}, T., {B{\'e}zard}, B., {Bushroe},
  M., {Combes}, M., {Cook}, D., {Coustenis}, A., {de Bergh}, C., {Dafoe},
  L.~E., {Doose}, L., {Dout{\'e}}, S., {Eibl}, A., {Engel}, S., {Gliem}, F.,
  {Grieger}, B., {Holso}, K., {Howington-Kraus}, E., {Karkoschka}, E.,
  {Keller}, H.~U., {Kirk}, R., {Kramm}, R., {K{\"u}ppers}, M., {Lanagan}, P.,
  {Lellouch}, E., {Lemmon}, M., {Lunine}, J., {McFarlane}, E., {Moores}, J.,
  {Prout}, G.~M., {Rizk}, B., {Rosiek}, M., {Rueffer}, P., {Schr{\"o}der},
  S.~E., {Schmitt}, B., {See}, C., {Smith}, P., {Soderblom}, L., {Thomas}, N.,
  {West}, R., Dec. 2005. {Rain, winds and haze during the Huygens probe's
  descent to Titan's surface}. \nat 438, 765--778.

\bibitem[{{Tomasko} et~al.(2002){Tomasko}, {Buchhauser}, {Bushroe}, {Dafoe},
  {Doose}, {Eibl}, {Fellows}, {Farlane}, {Prout}, {Pringle}, {Rizk}, {See},
  {Smith}, and {Tsetsenekos}}]{T02}
{Tomasko}, M.~G., {Buchhauser}, D., {Bushroe}, M., {Dafoe}, L.~E., {Doose},
  L.~R., {Eibl}, A., {Fellows}, C., {Farlane}, E.~M., {Prout}, G.~M.,
  {Pringle}, M.~J., {Rizk}, B., {See}, C., {Smith}, P.~H., {Tsetsenekos}, K.,
  Jul. 2002. {The Descent Imager/Spectral Radiometer (DISR) Experiment on the
  Huygens Entry Probe of Titan}. Space Science Reviews 104, 469--551.

\bibitem[{{Tomasko} et~al.(2008){Tomasko}, {Doose}, {Engel}, {Dafoe}, {West},
  {Lemmon}, {Karkoschka}, and {See}}]{T07}
{Tomasko}, M.~G., {Doose}, L., {Engel}, S., {Dafoe}, L.~E., {West}, R.,
  {Lemmon}, M., {Karkoschka}, E., {See}, C., Apr. 2008. {A model of Titan's
  aerosols based on measurements made inside the atmosphere}. \planss 56,
  669--707.

\bibitem[{{Towner} et~al.(2006){Towner}, {Garry}, {Lorenz}, {Hagermann},
  {Hathi}, {Svedhem}, {Clark}, {Leese}, and {Zarnecki}}]{TG06}
{Towner}, M.~C., {Garry}, J.~R.~C., {Lorenz}, R.~D., {Hagermann}, A., {Hathi},
  B., {Svedhem}, H., {Clark}, B.~C., {Leese}, M.~R., {Zarnecki}, J.~C., Dec.
  2006. {Physical properties of Titan's surface at the Huygens landing site
  from the Surface Science Package Acoustic Properties sensor (API-S)}. Icarus
  185, 457--465.

\bibitem[{{West} et~al.(2005){West}, {Brown}, {Salinas}, {Bouchez}, and
  {Roe}}]{W05}
{West}, R.~A., {Brown}, M.~E., {Salinas}, S.~V., {Bouchez}, A.~H., {Roe},
  H.~G., Aug. 2005. {No oceans on Titan from the absence of a near-infrared
  specular reflection}. \nat 436, 670--672.

\bibitem[{{Zarnecki} et~al.(2005){Zarnecki}, {Leese}, {Hathi}, {Ball},
  {Hagermann}, {Towner}, {Lorenz}, {McDonnell}, {Green}, {Patel}, {Ringrose},
  {Rosenberg}, {Atkinson}, {Paton}, {Banaszkiewicz}, {Clark}, {Ferri},
  {Fulchignoni}, {Ghafoor}, {Kargl}, {Svedhem}, {Delderfield}, {Grande},
  {Parker}, {Challenor}, and {Geake}}]{Z05}
{Zarnecki}, J.~C., {Leese}, M.~R., {Hathi}, B., {Ball}, A.~J., {Hagermann}, A.,
  {Towner}, M.~C., {Lorenz}, R.~D., {McDonnell}, J.~A.~M., {Green}, S.~F.,
  {Patel}, M.~R., {Ringrose}, T.~J., {Rosenberg}, P.~D., {Atkinson}, K.~R.,
  {Paton}, M.~D., {Banaszkiewicz}, M., {Clark}, B.~C., {Ferri}, F.,
  {Fulchignoni}, M., {Ghafoor}, N.~A.~L., {Kargl}, G., {Svedhem}, H.,
  {Delderfield}, J., {Grande}, M., {Parker}, D.~J., {Challenor}, P.~G.,
  {Geake}, J.~E., Dec. 2005. {A soft solid surface on Titan as revealed by the
  Huygens Surface Science Package}. \nat 438, 792--795.

\end{thebibliography}

\end{document}